\documentclass[fleqn,usenatbib]{mnras}

\usepackage{newtxtext,newtxmath}

\usepackage[T1]{fontenc}

\DeclareRobustCommand{\VAN}[3]{#2}
\let\VANthebibliography\thebibliography
\def\thebibliography{\DeclareRobustCommand{\VAN}[3]{##3}\VANthebibliography}

\usepackage{hyperref}
\usepackage{enumitem}
\usepackage{amsmath}

\usepackage{soul}
\usepackage{longtable}
\usepackage{pdflscape}
\usepackage{comment}
\usepackage{booktabs}
\usepackage{multirow}

\usepackage{graphicx}		
\usepackage{latexsym}		
\usepackage{bm}			
\usepackage[english]{babel}
\usepackage{color}

\title[LF at $z\sim7$--10]{Galaxy Luminosity Functions at $\mathbf{z\sim7}$--10 from Four Years of JWST Observations}

\author[Sun \& Yan]{
Bangzheng Sun,$^{1,2}$\thanks{E-mail: bangzheng.sun@missouri.edu}
Haojing Yan$^{1}$
\\
$^{1}$Department of Physics and Astronomy, University of Missouri-Columbia, 701 S College Ave., Columbia, MO 65201, USA \\
$^{2}$IPAC, California Institute of Technology, 1200 E California Blvd., Pasadena, CA 91125, USA
}

\date{Accepted XXX. Received YYY; in original form ZZZ}

\pubyear{\the\year{}}

\begin{document}
\label{firstpage}
\pagerange{\pageref{firstpage}--\pageref{lastpage}}
\maketitle

\begin{abstract}

We present galaxy luminosity functions at $z\approx 7.4$ and 10 based on a
comprehensive search for candidate galaxies at $6.4\lesssim z\lesssim11.3$ 
in JWST's public NIRCam data four years after its operation. Our work
utilises nine fields covering $\sim$1,045~arcmin$^2$ in total, including seven
blank fields and two galaxy cluster fields. Using the dropout technique, we 
selected 3,733 dropouts from the F090W passband and 2,396 dropouts from
the F115W passband, which constitute the largest samples of candidate galaxies
at $z\approx 7.4$ and $z\approx 10$ to date. 
Among these objects, 338 F090W-dropouts and 99 F115W-dropouts 
already have secure redshifts 
from JWST NIRSpec spectroscopy over four years, which we use to form two large 
spectroscopic subsamples to assess the magnitude-dependent contamination rates 
of their parent dropout samples. Our luminosity functions are based on the 
results after statistically correcting for both incompleteness and 
contamination. We find that the luminosity function at $z\approx 7.4$ can be
well described by a conventional Schechter function but with a faint-end slope
of $\alpha=-1.70\pm0.16$ that is significantly shallower than most of the
previous studies at $z\approx 7$--8. To verify this result, more imaging and
spectroscopy at $\sim$30~mag will be needed. The luminosity function at 
$z\approx 10$, on the other hand, can hardly be fitted by a Schechter function
but is more consistent with a power law in the form of
$\Phi(M)\propto 10^{-0.4(\beta+1)M}$, with the best-fit slope of 
$\beta=-2.36\pm0.16$. While such a behavior is largely driven by the high number
densities at $M_{\rm UV}\lesssim -21.8$~mag, it is supported by the currently 
existing spectroscopic confirmations. Further spectroscopy at the bright end
($\lesssim 26$~mag) will be needed to strengthen this result.


\end{abstract}




\section{Introduction} \label{sec:intro}

   For more than two decades, understanding galaxy formation and evolution 
during the first several million years after the Big Bang has been one of the 
most exciting and yet the most challenging subjects.
The Hubble Space Telescope (HST) used to play the central role in our 
studies of early galaxies, which gradually advanced the galaxy redshift 
frontier from $z\approx 6$ to $z\approx 10$
\citep[e.g.,][]{Yan2004b,Bunker2004, Bouwens08, McLure09, Bouwens10, Bouwens11, 
Yan2012, Oesch18, Bowler20}. 
The advent of the James Webb Space Telescope (JWST) has led to a major leap forward 
in this field. The spectroscopically confirmed redshift record for 
galaxies was quickly updated from $z=13.2$ \citep[][]{CurtisLake23, 
Robertson2023a} to $z=14.32$ \citep[][]{Carniani2024} and to $z=14.44$
\citep[][]{Naidu2026}. Meanwhile, it is now a consensus that the number of 
galaxies at $z\gtrsim 10$ is at least an order of magnitude higher than the
pre-JWST estimates, which is evident from the searches for high-redshift
(high-$z$) candidate galaxies by various groups \citep[e.g.,][]{Naidu22, 
Donnan23, Yan_highz_23, Finkelstein2023,Harikane23, Atek23, Austin23, Adams23,
Donnan24, McLeod2024, Gonzalez2025, Whitler2025}, some of which have pushed to
$z\approx 25$.

   Despite the rapid development at the redshift forefront, there are still
many critical pieces needed to be finished at $z\lesssim 14$ to form a new 
picture of early galaxy formation and evolution. Measuring the luminosity
functions of galaxies at $z>7$ is one of them. While there have been many
such measurements using the JWST data \citep[e.g., ][]{Bouwens2023, Donnan23, Harikane23_uvlf, Willott24, Adams23, Donnan24, McLeod2024, Whitler2025}, 
there are still obvious inconsistencies across the entire
luminosity range, which severely hamper the calibrations of new theoretical
models. 
In this work, we focus on the redshift range of $z\approx 7$--10, where the
JWST has gathered sufficient imaging and spectroscopic data after four years
of operation to allow for significant improvements. 

   To this end, we conducted a systematic search for candidate galaxies at 
two redshift ranges, $z=6.4$--8.4 (hereafter $z\approx 7.4$) and $z=8.4$--11.3
(hereafter $z\approx 10$), using the public JWST NIRCam data in nine fields 
totaling $\sim$1,045~arcmin$^2$. The combination of different spatial coverage 
and depths in these fields naturally forms a multi-tiered dataset. Most of these
fields are widely separated from each other, and therefore the impact of the 
so-called ``cosmic variance'' is minimized. We selected the candidate galaxies 
using the classic dropout method. In total, we have 3,733 and 2,396 candidates
at $z\approx 7.4$ and 10, respectively, which constitute the largest samples of 
candidate galaxies at these redshifts to date. As it turned out, 338 of the
F090W-dropouts and 99 of the F115W-dropouts have secure redshifts from the JWST 
NIRSpec spectroscopy, which afforded us an unprecedented opportunity to directly 
assess the contamination rates in our candidate samples. Based on these results, 
we derive luminosity functions at $z\approx 7.4$ and 10.0 and discuss the
implications.

     The structure of this paper is as follows. We describe the imaging data 
used in this work and the data reduction procedures in Section \ref{sec:data}, 
along with the photometry in Section \ref{sec:phot}. Section~\ref{sec:dropouts} 
presents the dropout selection, while Section~\ref{sec:incompleteness} 
describes how the selection incompleteness is quantified through simulations. In 
Section~\ref{sec:specz}, we present the spectroscopic sample and derive the 
magnitude-dependent contamination rates for corrections. 
Section~\ref{sec:lf} presents the derivation of the luminosity functions and
the evolution from $z\approx 10$ to 7.4.
We summarize our main results in Section~\ref{sec:summary}. 
All magnitudes quoted in this work are in the AB system. 
All coordinates are in the ICRS frame and Equinox 2000. 
We adopt a flat $\Lambda$CDM cosmology with $H_0=71$ km s$^{-1}$ Mpc$^{-1}$, $\Omega_m=0.27$, and $\Omega_\Lambda=0.73$. 

\section{Public NIRCam Data in Nine Fields}\label{sec:data}

Most of the NIRCam data used in this work are the same as the ones used in
\citet[][]{Yan_highz_23}, and here we have added those that became public
since then. The data spread over seven blank fields and two galaxy 
cluster fields, all of which have been observed in at least six broadband 
filters. We reduced these data on our own, following the procedures as described
in \citet[][]{Yan_highz_23} and \citet[][]{SY25_vbd}.
These data and the reduction are briefly described below 
and are summarized in Table~\ref{tab:allfields}. 

\subsection{Field descriptions}

\begin{table*}
    \centering
    \begin{tabular}{ccccccc} \hline
         Field & Ctr. R.A. & Ctr. Decl. & Area (arcmin$^2$) & Exposure (ks) & Astrometry & HST Ancillary \\ \hline
         \multicolumn{7}{c}{\textit{Blank Fields}} \\ \hline
         PRIMER/COSMOS & 150.12299 & 2.34764 & 144.67 & $\sim1$--6 & CANDELS & CANDELS \\
         PRIMER/UDS & 34.35004 & -5.20001 & 247.38 & $\sim1$--4 & CANDELS & CANDELS \\
         EGS & 215.00545 & 52.93451 & 91.02 & $\sim2$--5 & CANDELS & CANDELS \\
         HUDF-par2 & 53.24923 & -27.84753 & 18.06 & $\sim50$--130 & UDF12 & own reduction \\
         GOODS-S & 53.16444 & -27.78256 & 84.35 & $\sim6$--60 & CANDELS & CANDELS \\
         GOODS-N & 189.23220 & 62.23857 & 87.88 & $\sim3$--25 & CANDELS & CANDELS \\
         NEXUS-Wide & 268.46205 & 65.19901 & 309.04 & $\sim1$ & GAIA DR3 & / \\
         \hline
         \multicolumn{7}{c}{\textit{Cluster Fields}} \\ \hline
         Abell 2744 & 3.55273 & -30.38000 & 48.68 & $\sim2$--20 & GAIA DR3 & / \\
         Abell 370 & 39.98595 & -1.59699 & 14.12 & $\sim5$--45 & HFF & HFF \\
         \hline
    \end{tabular}
    \caption{Basic properties of the public JWST NIRCam data used in this study, 
    grouped into blank fields and cluster fields as separated in the table. 
    Ctr.\ R.A.\ and Ctr.\ Decl.\ give the central coordinates of the final 
    mosaics, and the area listed is the region within each field covered by at 
    least five broadband filters. All NIRCam data are reduced using the JWST 
    calibration pipeline version 1.20.1 under the calibration context of 
    {\tt jwst\_1464.pmap}. The per-band exposure times (in kiloseconds) are the
    representative values. The last two column list the external catalogue used 
    for astrometric calibration (and image alignment) and the HST ancillary data 
    used in this study, respectively.
    For NEXUS-Wide, no HST ancillary data exist at the time of this study, and 
    for Abell~2744, we chose not to use the available HST ancillary data because 
    the NIRCam F070W data in this field are sufficiently deep for the veto
    purpose.
    }
    \label{tab:allfields}
\end{table*}

\begin{table*}
\centering
\resizebox{0.99\textwidth}{!}{
\begin{tabular}{lcccccccc} 
\toprule
 & \multicolumn{8}{c}{Band} \\
\cmidrule(lr){2-9}
Field & F090W & F115W & F150W & F200W & F277W & F356W & F410M & F444W \\
\midrule
\multicolumn{8}{l}{\textit{Blank Fields}} \\ \midrule
\multirow{2}{*}{PRIMER/COSMOS}
 & 27.21 & 27.72 & 27.92 & 28.12 & 28.68 & 28.77 & 28.03 & 28.41 \\
 & 26.70 -- 27.53 & 27.19 -- 28.05 & 27.37 -- 28.27 & 27.57 -- 28.47 & 28.09 -- 29.06 & 28.22 -- 29.14 & 27.51 -- 28.40 & 27.87 -- 28.78 \\ 
\midrule
\multirow{2}{*}{PRIMER/UDS}
 & 27.45 & 27.48 & 27.73 & 27.91 & 28.42 & 28.47 & 27.73 & 28.07 \\
 & 27.02 -- 27.73 & 27.06 -- 27.77 & 27.31 -- 28.02 & 27.49 -- 28.20 & 28.01 -- 28.70 & 28.08 -- 28.74 & 27.35 -- 28.00 & 27.68 -- 28.34 \\
\midrule
\multirow{2}{*}{EGS}
 & 28.31 & 28.33 & 28.13 & 28.42 & 28.83 & 28.90 & 28.15 & 28.50 \\ 
 & 28.08 -- 28.53 & 27.66 -- 28.57 & 27.71 -- 28.35 & 28.24 -- 28.59 & 28.39 -- 29.00 & 28.31 -- 29.06 & 27.66 -- 28.28 & 28.35 -- 28.74 \\ 
\midrule
\multirow{2}{*}{HUDF-par2}
 & - & 29.51 & 29.38 & 29.67 & 30.17 & 29.87 & 29.86 & 29.99 \\ 
 & - & 29.16 -- 29.87 & 28.97 -- 29.73 & 28.89 -- 30.26 & 29.76 -- 30.41 & 29.44 -- 30.25 & 29.10 -- 29.96 & 29.50 -- 30.40 \\ 
\midrule
\multirow{2}{*}{GOODS-S}
 & 28.63 & 28.76 & 28.86 & 29.00 & 29.40 & 29.49 & 28.92 & 29.14 \\ 
 & 28.15 -- 29.76 & 28.35 -- 29.94 & 28.41 -- 30.02 & 28.58 -- 30.04 & 29.07 -- 30.67 & 29.17 -- 30.55 & 28.56 -- 30.08 & 28.69 -- 30.23 \\ 
\midrule
\multirow{2}{*}{GOODS-N}
 & 28.49 & 28.65 & 28.66 & 28.68 & 29.21 & 29.23 & 28.68 & 29.00 \\ 
 & 27.59 -- 28.87 & 27.90 -- 28.96 & 27.80 -- 28.99 & 27.96 -- 29.00 & 28.39 -- 29.49 & 28.53 -- 29.47 & 27.90 -- 29.02 & 28.18 -- 29.36 \\ 
\midrule 
\multirow{2}{*}{NEXUS-Wide}
 & 26.93 & 26.81 & 26.79 & 28.05 & - & 27.66 & - & 27.87 \\ 
 & 26.64 -- 27.18 & 26.57 -- 27.02 & 26.64 -- 26.97 & 27.79 -- 28.25 & - & 27.45 -- 27.77 & - & 27.63 -- 27.99 \\
 
\midrule
\multicolumn{8}{l}{\textit{Cluster Fields}} \\ \midrule
\multirow{2}{*}{Abell 2744}
 & 29.00 & 28.67 & 28.71 & 28.73 & 29.13 & 29.34 & 28.68 & 29.13 \\ 
 & 28.46 -- 29.37 & 27.54 -- 29.15 & 27.56 -- 29.04 & 27.71 -- 29.06 & 28.07 -- 29.46 & 28.11 -- 29.64 & 28.28 -- 28.94 & 28.24 -- 30.36 \\

\midrule
\multirow{2}{*}{Abell 370}
 & 28.19 & 28.97 & 29.10 & 29.15 & 29.46 & 29.40 & 28.07 & 28.83 \\ 
 & 27.82 -- 28.43 & 28.05 -- 29.20 & 26.23 -- 29.50 & 26.88 -- 29.53 & 27.11 -- 29.86 & 27.14 -- 29.79 & 25.87 -- 28.65 & 28.41 -- 28.99 \\

\bottomrule
\end{tabular}
}
\caption{$5\sigma$ limiting magnitudes in the main NIRCam bands for the fields 
in Table~\ref{tab:allfields}.
For each field and band, the first row gives the median $5\sigma$ limiting magnitude within $r = 0\farcs2$; the second row gives the
16th--84th percentile range. 
}
\label{tab:depths}
\end{table*}

\subsubsection{Blank Fields}

$\bullet$ {\bf PRIMER in COSMOS and UDS} are from the Public Release IMaging for 
Extragalactic Research (PRIMER, PID 1837; \citealt[][]{Dunlop21}) program, which
cover 144.67 and 247.38~arcmin$^2$ in the COSMOS field and the UDS field, 
respectively. 
The survey utilized eight NIRCam 
filters: F090W, F115W, F150W, and F200W in the short-wavelength channel (SW), 
and F277W, F356W, F410M, and F444W in the long-wavelength channel (LW), 
enabling the selection of both F090W- and F115W-dropouts. 
To assist the F090W-dropout selection, we incorporated the HST ACS data from 
the Cosmic Assembly Near-IR Deep Extragalactic Legacy Survey
\citep[CANDELS;][]{Grogin2011,Koekemoer2011} in both fields. 
The PRIMER program executed the NIRCam imaging as coordinated parallels to its 
primary MIRI observations; as a result, the NIRCam exposures are highly 
non-uniform across each field, 
averaging $\sim$1--6~ks per band in COSMOS and $\sim$1--4~ks per band in UDS. 
The median $5\sigma$ point-source depths (within $r=0\farcs2$) of the stacks 
range from 27.21 (F090W) to 28.77~mag (F356W) in COSMOS, and from 27.01 (F200W) 
to 28.47~mag (F356W) in UDS (see Table~\ref{tab:depths}). 

$\bullet$ {\bf EGS}, combining the Cosmic Evolution Early Release Science Survey 
(CEERS, PID 1345; \citealt[][]{Finkelstein2023}) and PID 2234 (PI E. Banados) 
programs, covers 91.02~arcmin$^2$. The combined mosaic has nine NIRCam bands: 
F090W, F115W, F150W, and F200W in the SW channel, and F277W, F356W, F410M, 
F444W, and F470N in the LW channel, allowing the selection of both F090W- and 
F115W-dropouts. We also utilized the HST/ACS data from the CANDELS survey to 
assist the F090W-dropout selection. As both programs did the NIRCam observations 
as the coordinated parallels to the primary MIRI observations, the exposures are 
non-uniform across the field and range from $\sim$2 to 5~ks per band. 
The median $5\sigma$ depths range from 28.13 (F150W) to 28.90~mag (F356W). 
The median depth of the narrow-band filter F470N is of $25.91$~mag, 
and this filter is not used in our analysis. 

$\bullet$ {\bf HUDF-par2} combines the data from three programs: the MIRI HUDF 
Deep Imaging Survey (PID 1283; PI G. Oestlin), the Next Generation Deep 
Extragalactic Exploratory Public Survey (NGDEEP, PID 2079; 
\citealt[][]{Finkelstein2021}), and PID 6511 (PI G. Oestlin). 
Although it covers only 18.06~arcmin$^2$, it is one of the deepest NIRCam fields 
to date,  with average exposures of $\sim$50--130~ks per band and reaching
$\sim$250~ks in the deepest regions in F115W. The field was observed in ten 
NIRCam bands: F115W, F150W, F182M, F200W, and F210M in the SW channel, and 
F277W, F300M, F356W, F410M, and F444W in the LW channel. The median $5\sigma$ 
depths range from 29.38 (F150W) to 30.17~mag (F277W). Due to the lack of 
F090W, our selection in this field is limited to F115W-dropouts. We also reduced 
the HST/ACS imaging of this field to assist the dropout selection. 
Astrometry for all products is tied to the products of 2012 Hubble Ultra Deep 
Field \citep[UDF12;][]{Ellis2013_udf12, Koekemoer2013_udf12}. 

$\bullet$ {\bf GOODS-S} combines the data from the ``deep'' and ``medium'' tiers 
of the JWST Advanced Deep Extragalactic Survey (JADES, PID 1180, 1210, 
1286, 1287; \citealt[][]{Eisenstein2026_jades-overview, Bunker24, Hainline2024, Rieke2024, 
Eisenstein2025_jades}) as well as those from the Parallel wide-Area NIRCam Observations to Reveal And Measure the Invisible Cosmos survey (PANORAMIC, PID 2514; 
\citealt[][]{Williams2021}), resulting in a single contiguous field covering 
84.35~arcmin$^2$. 
The field has nine NIRCam bands: F090W, F115W, F150W, F200W in the SW channel, 
and F277W, F335M, F356W, F410M, and F444W in the LW channel, 
enabling both F090W- and F115W-dropout selection. 
Exposure times are highly non-uniform across the combined mosaic, ranging 
from $\sim$6 to $\sim$60~ks per band, with the shallowest regions from
the PANORAMIC program.
The median $5\sigma$ depths range from 28.63 (F090W) to 29.49~mag (F356W). 
The F335M band has the median depth of 29.28~mag.
For the F090W-dropout selection, we also incorporated the HST/ACS data from 
the CANDELS survey. 

$\bullet$ {\bf GOODS-N} combines the data from JADES (PID 1181; 
\citealt[][]{Eisenstein2017_gdn}), First Reionization Epoch Spectroscopically 
Complete Observations (FRESCO, PID 1895; \citealt[][]{Oesch21_fresco}), and 
PANORAMIC (PID 2514; \citealt[][]{Williams2021}), which cover a total of 
87.88~arcmin$^2$. The field has eight NIRCam bands: F090W, F115W, F150W, and 
F200W in SW, and F277W, F356W, F410M, and F444W in LW, enabling the selection of 
both F090W- and F115W-dropouts. Combining the three programs makes it another 
wide and deep field, only slightly shallower than GOODS-S. The exposure time 
ranges from $\sim$3--25~ks in different bands. The median $5\sigma$ depths range
from 28.49 (F090W) to 29.23~mag (F356W). In addition, part of the field also 
has data in F070W and F335M, reaching median depths of 28.12~mag and 28.69~mag, 
respectively. For the F090W-dropout selection, we used the HST/ACS data from the 
CANDELS survey. 
 
$\bullet$ {\bf NEXUS-Wide} is the widest field in our work, and we used
the data from the reference epoch of the North ecliptic pole EXtragalactic 
Unified Survey (NEXUS, PID 5105, PI Shen; \citealt[][]{Shen2024_NEXUS}) covering 
309.04~arcmin$^2$. The images are in six NIRCam bands: F090W, F115W, F150W, and 
F200W in the SW channel, and F356W and F444W in the LW channel.
The median $5\sigma$ depths ranging from 26.79 (F150W) to 28.05~mag (F200W). 
Astrometry was tied to \textit{Gaia} DR3. In this field, we only selected 
F115W-dropouts but not F090W-dropouts because the selection of the latter would
not be reliable due to the lack of images to the bluer side of F090W as the
``veto'' images.

\subsubsection{Cluster Fields}

$\bullet$ {\bf Abell 2744} combines the Ultradeep NIRSpec and NIRCam 
Observations before the Epoch of Reionization (UNCOVER, PID 2561; 
\citealt{Bezanson2024_uncover}), the Director's Discretionary Time (DDT) program 
PID 2756 \citep[][]{Chen22}, the ``Medium Bands, Mega Science'' survey 
(PID 4111; \citealt[][]{Suess2024_pid4111}), and the coordinated NIRCam parallel 
imaging from the GLASS survey (PID 1324; \citealt[][]{True2022}). 
Together, these programs result in the deepest NIRCam cluster field, 
covering 48.68~arcmin$^2$. The images are in nine NIRCam 
bands: F070W, F090W, F115W, F150W, and F200W in the SW channel, and F277W, 
F356W, F410M, and F444W in the LW channel, 
allowing the selection of both F090W and F115W dropouts. 
We registered the astrometry to \textit{Gaia} DR3.
The total exposure times are $\sim$10--20~ks per band over most of the area, 
but the outskirts contributed only by PID 2756 are shallower ($\sim2$--4~ks). 
The median $5\sigma$ depth in F070W is 28.46~mag, while those in other bands
range from 28.67 (F115W) to 29.34~mag (F356W). We did not use the HST data 
 because the NIRCam F070W already covers the blue end.
In addition, the field is also covered by an extensive set of medium-band filters, which are listed below with the median $5\sigma$ depths:
F140M (27.71), F162M (27.77), 
F182M (28.15), F210M (28.06), F250M (27.92), F300M (28.37), F335M (28.50), 
F360M (28.50), F430M (27.77), F460M (27.53), and F480M (27.73).

$\bullet$ {\bf Abell 370} combines the data from the Canadian NIRISS Unbiased 
Cluster Survey (CANUCS; PID 1208, PI C. Willott; 
\citealt[][]{Sarrouh2026_canucs}), PID 3538 (PI E. Iani), 
PID 5324 (PI J. Pierel), the JWST Ultimate Medium-band Photometric Survey 
(JUMPS; PID 5890, PI C. Withers), and the Vast Exploration for Nascent, 
Unexplored Sources program (VENUS; PID 6882, PI S. Fujimoto). 
The field covers 14.12~arcmin$^2$ and was observed in nine NIRCam bands: F070W,
F090W, F115W, F150W, and F200W in the SW channel, and F277W, F356W, F410M, and 
F444W in the LW channel, also allowing the selection of both F090W- and 
F115W-dropouts. The astrometric registration was tied to the Hubble Frontier 
Fields \citep[HFF;][]{Lotz2017}. The HFF HST/ACS data are also used to assist 
the selection of the F090W-dropouts. The median $5\sigma$ depths range from 
28.07 (F410M) to 29.46~mag (F277W), with substantial scatter across the field 
(e.g., the F150W and F200W 16th--84th percentile ranges span 26.23--29.50 
and 26.88--29.53~mag, respectively) reflecting the highly non-uniform coverage 
within this field. The field was also observed in multiple medium bands: F182M, 
F210M, F300M, F335M, F360M, F430M, F460M, and F480M, reaching median depths of 
27.84, 27.50, 27.16, 27.56, 28.69, 28.15, 27.28, and 27.60~mag, respectively.



\subsection{Data Reduction}

We reduced the aforementioned NIRCam data in a uniform way. The procedures 
began with the ``uncal'' products, which are individual exposures from the 
standard JWST data reduction pipeline 
\footnote{\href{https://github.com/spacetelescope/jwst}{https://github.com/spacetelescope/jwst}}
after the ``Level 1b'' processing and are available at the Mikulski Archive for 
Space Telescopes 
(MAST\footnote{\href{https://mast.stsci.edu/portal/Mashup/Clients/Mast/Portal.html}{https://mast.stsci.edu/portal/Mashup/Clients/Mast/Portal.html}}). 
We then used the version {\tt jwst\_1.20.1} pipeline and  
the reference ``context''\footnote{\href{https://jwst-crds.stsci.edu/}{https://jwst-crds.stsci.edu/}} {\tt jwst\_1464.pmap} for the remaining steps.


We first ran ``Stage 1'' of the pipeline, {\tt calwebb\_detector1},
to the ``uncal.fits'' images. This stage applies detector-level corrections and 
outputs count-rate ``rate.fits" images in the unit of 
counts per second. We adopted the default parameters except for those 
involving the detection and removal of ``snowballs'' (large cosmic-ray 
events); specifically, {\tt jump.expand\_large\_events} was set to ``True'' 
and {\tt jump.expand\_factor} was set to 1.5 so that the neighbouring pixels 
of the detected snowballs were also flagged. The vast majority of the SW 
count-rate images showed notable differences in the baseline (bias) levels 
in between the SW detectors. To mitigate this problem, we subtracted the 
median background count-rate value from any individual exposure in the 
SW channel. To estimate the background, each image was segmented into blocks 
of $128\times128$ pixels with sources masked and was applied a $3\times3$ 
median filter; the global median value of the filtered image was then taken 
as the background. The images in the SW detectors A3, A4, B3, and B4 
are known to have the defects called ``wisps''. We 
removed such defects by subtracting the corresponding templates
\footnote{\href{https://jwst-docs.stsci.edu/jwst-near-infrared-camera/nircam-instrument-features-and-caveats/nircam-claws-and-wisps}{https://jwst-docs.stsci.edu/jwst-near-infrared-camera/nircam-instrument-features-and-caveats/nircam-claws-and-wisps}}. 

The resultant images were then processed through ``Stage 2'' of the  pipeline, 
{\tt calwebb\_image2}. 
This stage applies flat-fielding and flux calibration,
and outputs the ``cal.fits'' images in units of MJy/sr. 
We adopted all default parameters set in the pipeline for this step. 
The SW images are known to be affected by the ``$1/f$'' readout noise pattern, 
which we corrected using the algorithm ``image1overf''\footnote{\href{https://github.com/chriswillott/jwst}{https://github.com/chriswillott/jwst}} 
on a per-amplifier basis along both rows and columns of the image. 
We also applied the same correction to the LW images when a clear ``$1/f$'' noise pattern was visually apparent. 

The image alignment (and hence the astrometric registration) was performed using 
the {\tt TweakRegStep} functionality of the pipeline but with some modifications.
Instead of using the default source detection algorithm, 
we supplied our custom source catalogues for each ``cal.fits'' frame, 
which were then used by {\tt TweakRegStep} to align to the reference catalog. 

In the last step, we used ``Stage 3'' of the pipeline, {\tt calwebb\_image3}, to
generate the final mosaic images. Since the ``cal.fits'' images still show 
notable background variations, we subtracted a smooth, two-dimensional (2D) 
background from each image prior to mosaicking. 
This 2D background was derived following the same procedure described above for 
the count-rate images, except that a median-filtered 2D background was used 
instead of a single value. The background-subtracted  ``cal.fits'' images were 
then passed to {\tt calwebb\_image3} to produce the final mosaics in all bands. 
All data reduced and used in this study have a pixel scale of 0\farcs06 
(60~mas). 

    Some images suffer from contaminants of unknown origins, which 
required special treatments as described below.

\noindent $\bullet$ PRIMER/UDS: there are linear structures spanning substantial areas in the eastern part of the field. 
These features consistently appear at nearly the same celestial positions across observations spanning $\sim$6 months, suggesting that these are not
detector- or exposure-specific artifacts.
We suspect that these could be caused by an extremely bright object outside of
the field. These features cannot be removed and are left in the final
mosaics.

\noindent $\bullet$ EGS: there are prominent bright edges in the F200W images in 
Observation 52. We flagged these edges as {\tt DO\_NOT\_USE} in their 
respective data quality ({\tt DQ}) arrays so that they were not used 
during the mosaicking process. 

    
\noindent $\bullet$ GOODS-N: the shapes 
of the wisps in F115W have varied from the latest version of the officially 
available templates, and a direct subtraction of the templates did not work. To 
mitigate this problem, we constructed specific templates using these data on a 
per-detector basis: 
for each affected detector, we collected all individual exposures, 
masked the detected sources in each exposure, 
and combined the source-masked exposures using the biweight location statistics from \texttt{astropy.stats.biweight\_location}. 
The resulting per-detector templates were then subtracted from the corresponding single exposures. 

\section{Photometry for Dropout Selection}\label{sec:phot}


  We used SExtractor \citep{Bertin1996} in dual-image mode for source 
extraction and matched-aperture photometry. For all fields, we adopt the F356W
image as the detection image for several reasons: 
(1) The F356W image is one of the deepest across all bands. 
(2) The point spread function (PSF) in this band is sufficiently large, 
ensuring that the apertures determined within this band include all source 
fluxes in any bluer bands. 
(3) The F356W image has a notably cleaner background than those taken in 
the SW bands. We adopted the {\tt MAG\_ISO} magnitudes from SExtractor to 
minimize background noise. 

   We applied a $5\times5$ Gaussian convolution filter with a full width at 
half maximum (FWHM) of 2 pixels. The detection and analysis thresholds were set 
to 1.0. Only sources that had a signal-to-noise ratio (S/N) of at least 5 and 
ISOAREA of at least 10 pixels in the F356W detection band were retained.

   The photometric errors were estimated using the ``root mean square'' maps 
that we derived using the {\tt astroRMS} tool.\footnote{Courtesy of M. Mechtley; see \href{https://github.com/mmechtley/astroRMS}{https://github.com/mmechtley/astroRMS}. }
Specifically, the rms map of a given band was calculated using the weight 
extension (``WHT") from the direct pipeline product by $rms=A/\sqrt{\rm WHT}$, 
where $A$ is the scaling factor that is determined by considering 
the artificial suppression of noise due to pixel resampling. 
We modified the original {\tt astroRMS} code for better source masking. 

   Hereafter, we denote $m_{090}$, $m_{115}$, $m_{150}$, $m_{200}$, $m_{277}$, 
$m_{335}$, $m_{356}$, $m_{410}$, $m_{444}$ as the magnitudes in JWST/NIRCam 
F090W, F115W, F150W, F200W, F277W, F335M, F356W, F410M, and F444W bands, 
respectively, and $m_{606}$ and $m_{814}$ 
as the magnitudes in HST/ACS F606W and F814W bands, respectively. 

\section{Dropout Selection}\label{sec:dropouts}

   Our method of selecting F090W- and F115W-dropouts relies on the absorption 
associated with Lyman-$\alpha$ (Ly$\alpha$, rest-frame 1216~\AA) and the Lyman 
limit (rest-frame 912~\AA). This absorption, which is caused by discrete neutral 
hydrogen (H\,\textsc{i}) clouds along the line of sight, greatly suppresses
the flux at wavelengths bluer than rest-frame 1216~\AA\ in the spectral energy
distributions (SEDs) of the 
source, creating the so-called ``Lyman break'' signature.
Consequently, when such a source is observed in two adjacent bands, one
straddling the break and the other sampling the redder wavelengths, the source 
would be very faint or even undetected in the bluer band (the ``dropout 
band'') but would remain unaffected in the redder band (the ``shift-in band''). 
In our selection, we used the photometry done on the non-convolved images as
discussed in Section 3.1 to minimize the problem of blending with neighbours.


The dropouts were selected using the selection criteria following the methodology in \cite{Yan2023a}, which are detailed below.

$\bullet$ {\bf SNR $\geq5$ in the shift-in band.} 
This criterion ensures a robust detection in the shift-in band and therefore
the reliability of the measured dropout amplitude. 
The shift-in bands for F090W and F115W dropouts are F115W and F150W, respectively, for all fields. 

$\bullet$ {\bf Dropout amplitude $\geq$0.8~mag.} 
The dropout amplitude is defined as the colour difference between the dropout and 
the shift-in bands. For sources with S/N~$<2$ in the dropout band, 
we substitute the magnitude with the 2$\sigma$ upper limit measured within a 
circular aperture of $r=\sqrt{\rm ISOAREA_{F356W}/\pi}$ centred on the source. 
The threshold of 0.8~mag is adopted because a break shifted halfway out of the 
dropout band would make the object $\sim$0.75~mag fainter than no break is 
present (assuming a flat SED in $f_\nu$).

$\bullet$ {\bf SNR $\geq5$ in at least one additional band redder than the shift-in band (excluding F356W). } 
Recall that our detection criteria already require S/N~$\geq$5 in the detection band (F356W). 
Here we further require S/N~$\geq5$ in at least one additional band, 
so that each candidates meets high-S/N conditions in at least three bands, 
further strengthening the reliability of the detection. 

$\bullet$ {\bf Non-detections in all bands bluer than the dropout band (``veto" 
bands).} Since the Lyman break removes essentially all flux bluer of it when at 
$z\gtrsim 6$, we require S/N~$<2$, i.e., a non-detection, in all veto bands. 
Because the depth of the veto bands may differ significantly from that of the 
dropout and the shift-in bands, a candidate could mistakenly bypass the veto due 
to this depth disparity. Therefore, we impose an additional requirement that at 
least one veto band should be sufficiently deep, i.e., its $2\sigma$ depth 
satisfies $\lim_{\rm veto} - m_{\rm shift\text{-}in} \geq 0.8$, which ensures
that the veto band image is deep enough to support the non-detection. 

    The candidates that survived the above criteria were all visually inspected. 
This inspection allowed us to further reject the false detections due to image 
artifacts and to confirm that each candidate is genuinely undetected in all veto
bands. 

   Some of the surviving candidates are very bright, with the extreme ones even
approaching $m_{356}\lesssim22.0$~mag. \citet[][]{SY25_vbd} discussed the nature
of very bright dropouts that have $m_{356}<25.1$~mag, which would correspond to 
$M_{\rm UV}<-22.0$ at $z=7.4$. The basic conclusion is that a few of them could
be at $z\geq 6$ (see \citealt[][]{Haro2023} for one such object 
spectroscopically confirmed at $z=8.679$) but the majority should be low-redshift
``interlopers''. In this current work, we take a conservative approach and
exclude those that have $m_{356}<25.1$~mag.

   After this magnitude cut, our sample has 3,733 F090W dropouts and 2,396 F115W 
dropouts retained. The details of their distributions in the nine fields are 
summarized in Table~\ref{tab:dropouts}. 


\begin{table*}
\resizebox{0.95\textwidth}{!}{
\begin{tabular}{lcccccccccccccc}\hline
Field/Type & COSMOS & UDS & EGS & HUDF-par2 & GOODS-S & GOODS-N & NEXUS-Wide & Abell 2744 & Abell 370 & Total \\ \hline
F090W-dropouts & 553 & 472 & 784 & - & 709 & 673 & - & 492 & 50 & 3733 \\
\hline
F115W-dropouts & 616 & 569 & 101 & 143 & 402 & 168 & 254 & 74 & 69 & 2396 \\ 
\hline
\end{tabular}
}
\caption{Number of F090W- and F115W-dropouts retained in each field. 
}
\label{tab:dropouts}
\end{table*}

\begin{figure*}
\centering
\includegraphics[width=0.95\textwidth]{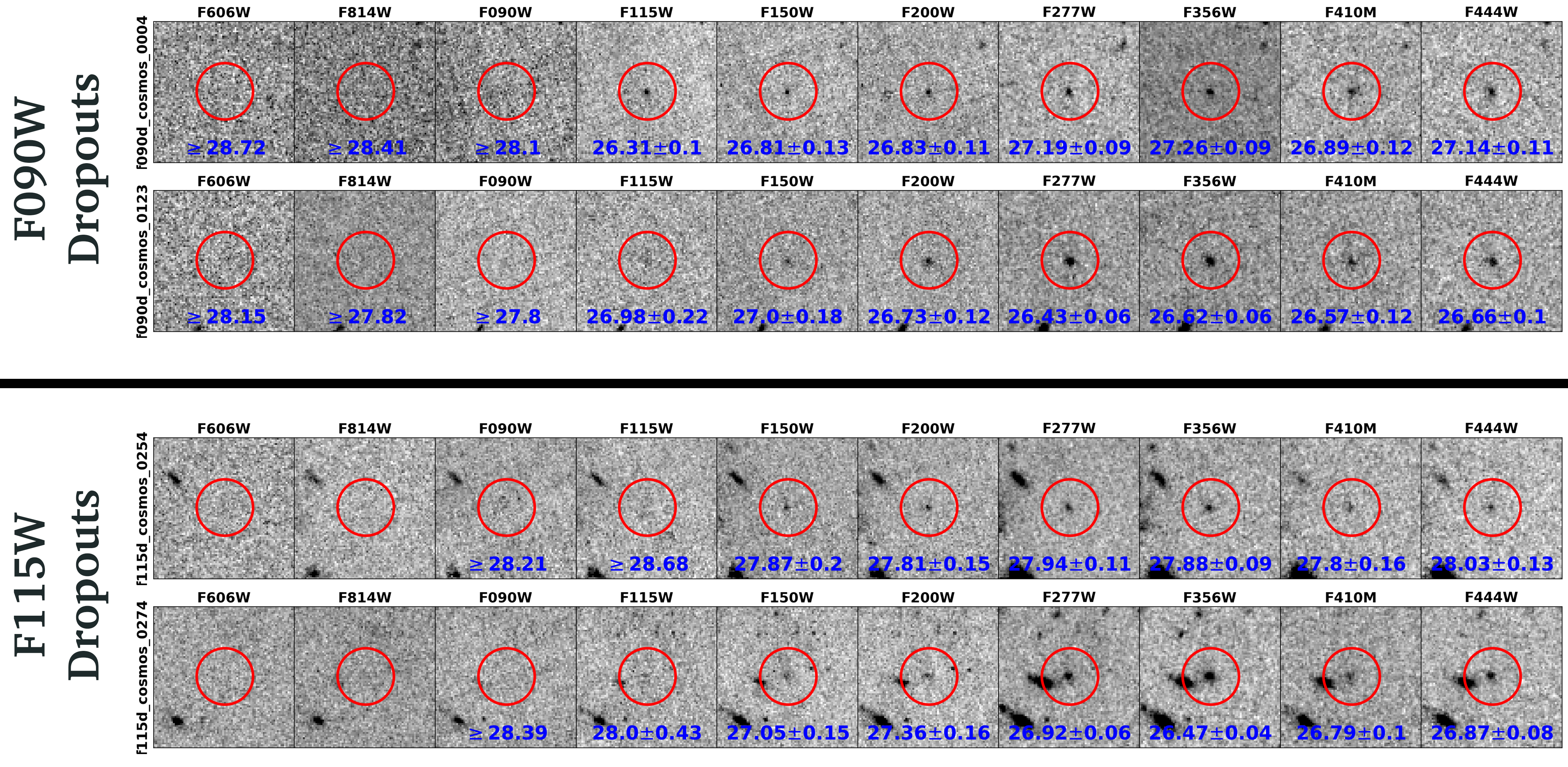}
\caption{Examples of selected dropouts in F090W (top two panels) and F115W (bottom two panels), respectively. 
All image stamps are $4\farcs8\times4\farcs8$ in size 
and the dropouts are enclosed in red circles with $r=0\farcs96$. 
Photometry and photometric errors in each band are given in blue text at the 
bottom of each image stamp; 
when the object is a non-detection in a certain band, the $2\sigma$ magnitude 
limit within $r=0\farcs2$ is provided instead. 
}
\label{fig:dropoutsample}
\end{figure*}

\section{Incompleteness Correction}\label{sec:incompleteness}

  Dropout selection is subject to incompleteness in that source detection can 
never be 100\% complete when approaching the S/N thresholds. 
If this is left uncorrected, it would bias the number density estimates, 
particularly near the faint end where this effect is the most severe. 
Therefore, we quantify the incompleteness and correct it by simulations.

   As the vast majority of our objects cannot be well described by analytic 
models, we opt to do simulations using real dropout images on a per-dropout
basis. Using real dropouts also circumvent the common problem in the simulations based on analytic models, which have to assume some certain SED behaviors.
Our approach is to place
the image cutouts of each dropout at random locations across the field and 
measure the recovery rate. For each dropout, we make $20\times20$-pixel 
cutouts in all bands centred on the source and then ``paste'' them to a large
number of random locations in the images of the corresponding bands. The
``pasting'' is done by replacing the pixel values in a $20\times20$-pixel area 
around the random locations with those of the cutout. 
This replacement is applied only to the science images,
while the corresponding RMS maps are left unaltered. 
For the large-area fields
($\geq 50$~arcmin$^2$; see Table~\ref{tab:allfields}), we generate 15{,}000
random locations; for all other fields, we generate 8{,}000. 

   The depth variation due to the non-uniform exposure times across the field
needs to be accounted for in the simulations. For example, if the original 
location of a source lies in a deep region while one of the simulated positions 
falls in a shallower one, the photometric error at that simulated position is 
significantly larger, which would artificially lower the recovery rate, i.e., 
the true recovery rate would be underestimated. To avoid this possible bias, 
we require the exposure time at each simulated position to be 
within 20\% of the exposure time at the original location where the dropout is 
found. 

    We then perform photometry to determine the fraction of usable simulated 
objects that can be recovered. Using the same {\tt SExtractor} configuration 
described in Section~\ref{sec:phot}, we retain simulated objects satisfying: 
(1) S/N~$\geq5$ and ISOAREA~$\geq10$ in the detection band (F356W); 
(2) S/N~$\geq5$ in the shift-in band; 
(3) S/N~$\geq5$ in one or more bands redder than the shift-in band (excluding
F356W).  
The recovery rate, $p_r$, is defined as the ratio of the recovered objects 
(i.e., satisfying all three conditions above) to the total number of simulated 
objects.

    Note that these conditions match the dropout selection criteria described 
earlier, with two exceptions: 
(1) The dropout amplitude requirement of $\geq0.8$~mag is omitted intentionally
because applying it here would not be appropriate due to the fact that we use 
the images of real objects (the retained dropouts) for the simulation. If the 
photometric error distributions are Gaussian, applying a break amplitude
criterion means that the number of objects being scattered above the criterion 
(and hence being selected) due to photometric errors is the same as the number 
of objects being scattered below the criterion (and hence being missed) for the 
same reason. In a sense, this is to say that the incompleteness due to applying 
the break amplitude criterion is zero because of the cancellation. As our 
simulation is done on a per-object basis using every retained dropout, we cannot 
simulate the ones being missed by our sample due to photometric errors. 
Therefore, if the break amplitude criterion were to be applied, the recovery 
rate would be highly underestimated, which would then lead to an erroneous, 
highly overestimated incompleteness correction.
(2) We do not need to consider non-detections in the veto bands because all the retained dropouts have been visually examined to satisfy the veto condition.

\section{Spectroscopic Identifications and Contamination Corrections}\label{sec:specz}

   There have been some large JWST NIRSpec spectroscopic programs in most 
of the fields used in this work (except HUDF-par2), and a significant number of 
our targets have been observed. This affords us a unique opportunity to 
assess the contamination rate of our sample by spectroscopy and to make 
corrections accordingly.

\subsection{Spectroscopic Sample}
 
We cross-match our dropout sample to the ``v4.4'' spectroscopic redshift catalogues hosted at the DAWN JWST Archive (DJA) \footnote{The catalog is retrieved from \href{https://s3.amazonaws.com/msaexp-nirspec/extractions/nirspec_public_v4.4.html}{https://s3.amazonaws.com/msaexp-nirspec/extractions/nirspec\_public\_v4.4.html}; also see \href{https://dawn-cph.github.io/dja/spectroscopy/nirspec}{https://dawn-cph.github.io/dja/spectroscopy/nirspec} for their reduction procedures, included programs, etc. }, 
using a matching radius of $r=0\farcs3$. This yields 437 matches across all 
fields. The spectroscopic redshifts we used are from a total number of 35 JWST 
programs: PIDs 1180, 1181, 1208, 1210, 1211, 1212,
1213, 1214, 1215, 1286, 1287, 1324, 1345, 1671, 1871, 1879, 1914,
2198, 2561, 2674, 2750, 2756, 3073, 3215, 3543, 3567, 4106, 4233,
4287, 4762, 5105, 5224, 6368, 6541, and 6585.
Based on the DJA ranking scheme, in which each spectroscopic redshift ($z_{\rm spec}$) is 
assigned a quality rank from 0 (unreliable) to 3 (secure), we retain only 
the sources with the rank of 2 or 3, i.e., those with reliable spectroscopic 
redshifts; matches with the rank of 0 or 1 are excluded from further analysis, as
they cannot be used to determine whether a candidate is a genuine high-$z$
galaxy or a low-$z$ interloper.

\begin{table}
\centering
\begin{tabular}{lccc}
\hline
Field & Drop-out band & High-$z$ & Interlopers \\ \hline 
COSMOS & F090W & 9 & 6 \\
       & F115W & 6 & 6 \\
       \hline 
UDS & F090W & 46 & 8 \\ 
    & F115W & 9 & 16 \\ 
    \hline 
EGS & F090W & 91 & 20 \\ 
    & F115W & 14 & 3 \\ 
    \hline 
GOODS-S & F090W & 52 & 10 \\
        & F115W & 17 & 2 \\ 
        \hline 
GOODS-N & F090W & 50 & 2 \\ 
        & F115W & 11 & 1 \\ 
        \hline 
NEXUS-Wide & F115W & 0 & 2 \\ 
           \hline 
Abell~2744 & F090W & 33 & 9 \\ 
           & F115W & 10 & 0 \\ 
           \hline 
Abell~370 & F090W & 2 & 0 \\ 
          & F115W & 0 & 2 \\ 
\hline
\end{tabular}
\caption{
Statistics of spectroscopic identifications in eight fields. 
``High-$z$'' indicates the number of selected dropouts in our sample that
are confirmed to be genuine high-redshift galaxies via spectroscopy;
``Interlopers'' is the number of confirmed low-redshift interlopers. 
HUDF-par2 is not included here as no NIRSpec programs cover this field. 
}
\label{tab:specz_stats}
\end{table}

    Table~\ref{tab:specz_stats} summarizes the statistics in each field. A full 
list of spectroscopically identified candidates is included in 
Appendix~\ref{sec:all_spec_z}. In total, the spectroscopically identified 
F090W-dropout sample contains 338 objects, of which 283 are confirmed at 
$z_{\rm spec}\geq 6.4$ and 55 are lower-$z$ interlopers. 
The spectroscopically identified F115W-dropout sample contains 99 objects, 
including 67 confirmed at $z_{\rm spec}\geq 8.4$ and 32 low-$z$ interlopers. 
These statistics indicate that the overall interloper fractions are 16.3\% and 
32.3\% for the F090W- and F115W-dropout samples, respectively. However, a
more careful treatment is necessary, as described below.

\subsection{Magnitude-dependent Contamination Rates}\label{sec:cont_rate}

Within each magnitude bin $k$, we estimate the contamination fraction as $f_{\rm ctm, \it k}=N_{\rm interloper,\it k}/N_{\rm spec,\it k}$, 
where $N_{\rm interloper,\it k}$ is the number spectroscopically confirmed low-redshift interlopers 
and $N_{\rm spec,\it k}$ is the total number of objects within the magnitude bin that have confirmed $z_{\rm spec}$. 
The statistical uncertainty of the contamination fraction in each bin is estimated by 
$\sigma_{k}=[f_{\rm ctm,\it k}(1-f_{\rm ctm, \it k})/N_{\rm spec,\it k}]^{1/2}$. 
We note that the contamination fraction could be unreliable if $N_{\rm spec,\it k}$ is small in a certain magnitude bin, 
and therefore, we only apply such a correction when $N_{\rm spec,\it k}\geq 3$. 

\begin{figure*}
    \includegraphics[width=0.99\textwidth]{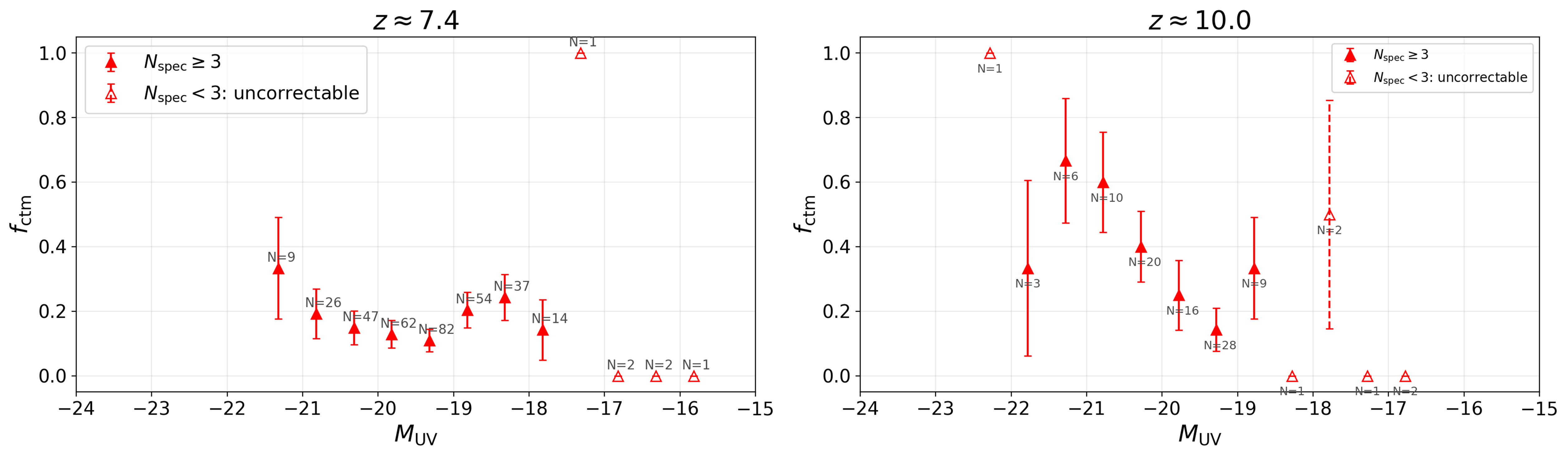}
    \caption{
    Magnitude-dependent contamination fractions for the F090W- and 
    F115W-dropout samples, respectively. The filled triangles indicate the 
    magnitude bins having at least three spectroscopically identified sources 
    ($N_{\rm spec}\geq3$), where the contamination correction is applied. The 
    error bars are based on the Poissonian noise. Open triangles, on the other 
    hand, indicate bins with $N_{\rm spec}<3$, for which the contamination 
    fraction is poorly constrained and no such correction can be made. 
    The number of $N_{\rm spec}$ in each bin is labeled. 
    }
    \label{fig:cont}
\end{figure*}

The magnitude-dependent contamination fractions are shown in
Figure~\ref{fig:cont}. For both samples, the trend is not flat: the 
contamination rate reaches the lowest at around $M_{\rm UV}=-20$ to $-19$~mag 
and increases to both ends. This is more obvious in the F115W-dropout sample.
Therefore, adopting a single contamination fraction for the entire dropout sample is inappropriate, especially for the F115W dropout sample. 

\section{Luminosity Functions}\label{sec:lf}

\subsection{Effective selection areas} \label{subsec:aeff}

In most fields, the effective selection area of a given candidate, i.e., 
the total area of sky within a field over which that candidate could have been 
selected, is considerably smaller than the total field area. 
This is mainly because of the highly varying exposure times across the field, 
particularly for the wide fields observed as coordinated NIRCam parallels, 
such as COSMOS, UDS, and EGS. 
For example, a faint dropout found in the deepest region of a field 
might fail to pass our S/N criteria had it been located
in a shallower region of the same field. Therefore,
the effective area must be calculated on a per-dropout basis. 

  For this purpose, we first generate the 2$\sigma$ and 5$\sigma$ ``depth maps'' 
for \emph{each dropout} based on the RMS maps: on a per-pixel basis, we compute
the depths (in magnitudes) within a radius of 
$r=\sqrt{{\rm ISOAREA_{F356W}}/\pi}$, 
consistent with the radius adopted for estimating the detection upper limits 
when constructing the SED. Based on these depth maps, we then compute the 
effective area for each candidate as follows: 

$\bullet$ {\bf Step 1}: 
determine $A_1$, the total area over which the $5\sigma$ depth in F356W (the detection band) 
is at least as deep as the magnitude of the candidate in this band. 
This ensures the source can be detected with SNR~$\geq 5$ in F356W. 

$\bullet$ {\bf Step 2}: 
within $A_1$, determine $A_{12}$, 
the total area over which the $5\sigma$ depth in the shift-in band is at least as deep as 
the magnitude of the candidate in this band. 
This ensures that the source can be detected with SNR~$\geq 5$ in the shift-in band. 

$\bullet$ {\bf Step 3}: 
within $A_{12}$, determine $A_{123}$, 
the total area over which the $2\sigma$ limit in the dropout band is at least 0.8~mag 
deeper than the magnitude of the candidate in the shift-in band, 
i.e., $\lim_{2\sigma,{\rm dropout}} - m_{\rm shift\text{-}in}\geq 0.8$~mag. 

   We then adopt $A_{123}$ as the effective area, $A_{\rm eff}$, for the given
candidate. For example, consider an F115W dropout that shifts in at F150W, 
with $m_{150}=28.50$ and $m_{356}=27.00$~mag. 
The effective area for this source, $A_{\rm eff}$, 
is the area covered by all pixels in the depth maps that simultaneously satisfy 
${\rm lim}_{5\sigma,356}\geq27.00$~mag, 
${\rm lim}_{5\sigma,150}\geq28.50$~mag, 
and ${\rm lim}_{2\sigma,115}\geq29.30$~mag 
(i.e., ${\rm lim}_{2\sigma,115}-m_{150}\geq0.8$~mag). 

    In the lensing cluster fields, we must also consider the magnification 
effects. For ease of processing, we de-magnify the fields and the brightness of
the dropouts by applying proper lensing models such that the simulations can be 
done in the same way as in the blank fields. 
We adopt the lens model from the fourth data release of the UNCOVER 
program \citep[][]{Furtak2023,Price2025} for the Abell~2744 field, and
the Hubble Frontier Fields 
v4 lens model of Sharon\footnote{\href{https://archive.stsci.edu/pub/hlsp/frontier/abell370/models/sharon/v4/}{https://archive.stsci.edu/pub/hlsp/frontier/abell370/models/sharon/v4/}} 
for the Abell~370 field. 
In Abell~2744, the model provides the $z=10$ magnification map that we can
directly use for the F115W dropouts. For the F090W dropouts, we estimate the magnification map at $z=7.4$ by averaging the provided maps at $z=6$ and $z=8$. 
In Abell~370, we use the tool provided by the HFF program 
to derive the magnification maps at $z=7.4$ and $z=10$, respectively. 
The field depths are deepened by applying the magnification factor on a 
per-pixel basis, and the magnitudes of the dropouts are replaced by their 
de-magnified values.

\subsection{Stepwise Luminosity Functions}

We derive the luminosity functions of galaxies at the nominal redshifts 
of the F090W and F115W dropouts, $z\approx 7.4$ and $z\approx10$, respectively. 
For each dropout, we denote $p_r$ as the recovery rate from the incompleteness correction 
described in Section~\ref{sec:incompleteness}, 
and $A_{\rm eff}$ as the effective area described in Section~\ref{subsec:aeff}. 
The contribution of an individual dropout $i$ to the surface density is then 
$\rho_i=(p_{r,i}\times A_{{\rm eff},i})^{-1}$, 
with the corresponding error estimated from Poisson statistics as $\Delta \rho_i=\rho_i$. 
For all dropouts in a given field $f$, 
the total surface density is $\rho_f=\sum_i \rho_i$, 
with total error $\Delta \rho_f=\sqrt{\sum_i(\Delta \rho_i)^2}$. 
The surface density combining all fields is then computed as the weighted average, 
$\rho=(\sum_f \rho_f w_f)/\sum_f w_f$, 
where the weights are $w_f=(\Delta \rho_f)^{-2}$. 

The luminosity function gives the number density per magnitude bin 
as a function of the absolute magnitude $M_{\rm UV}$, 
which we calculate from $m_{200}$ at $z\approx 7.4$ and from $m_{277}$ at $z\approx10$, 
since these bands correspond to similar rest-frame UV wavelengths of 
$\sim$2370~\AA\ and $\sim$2520~\AA, respectively. 
The number density is obtained by dividing the surface density by the comoving volume 
per unit area at the corresponding redshift. 
The redshift ranges spanned by the F090W and F115W dropouts 
are $z=6.4$--8.4 and $z=8.4$--11.3, respectively, 
over which the comoving volumes per unit area are 
$4.53\times10^{3}\ {\rm Mpc}^{3}\,{\rm arcmin}^{-2}$ and 
$5.19\times10^{3}\ {\rm Mpc}^{3}\,{\rm arcmin}^{-2}$, respectively. 

Finally, we correct for the magnitude-dependent contamination rates described 
in Section~\ref{sec:cont_rate}. We denote the uncorrected luminosity 
function estimate in magnitude bin $k$ as $\Phi_{\rm raw, \it k}$; 
for the ``correctable'' bins, the corrected number density is 
$\Phi_{\rm corr,\it k}=(1-f_{\rm ctm,\it k})\Phi_{\rm raw, \it k}$. 
Again, as stated in Section~\ref{sec:cont_rate}, for the bins containing 
less than 3 spectroscopically identified objects, we do not apply this 
correction, and we also exclude these data points when analyzing the behavior 
of luminosity functions. These results are 
listed in Table~\ref{tab:lf}. 

\begin{figure*}
\centering
\includegraphics[width=0.8\textwidth]{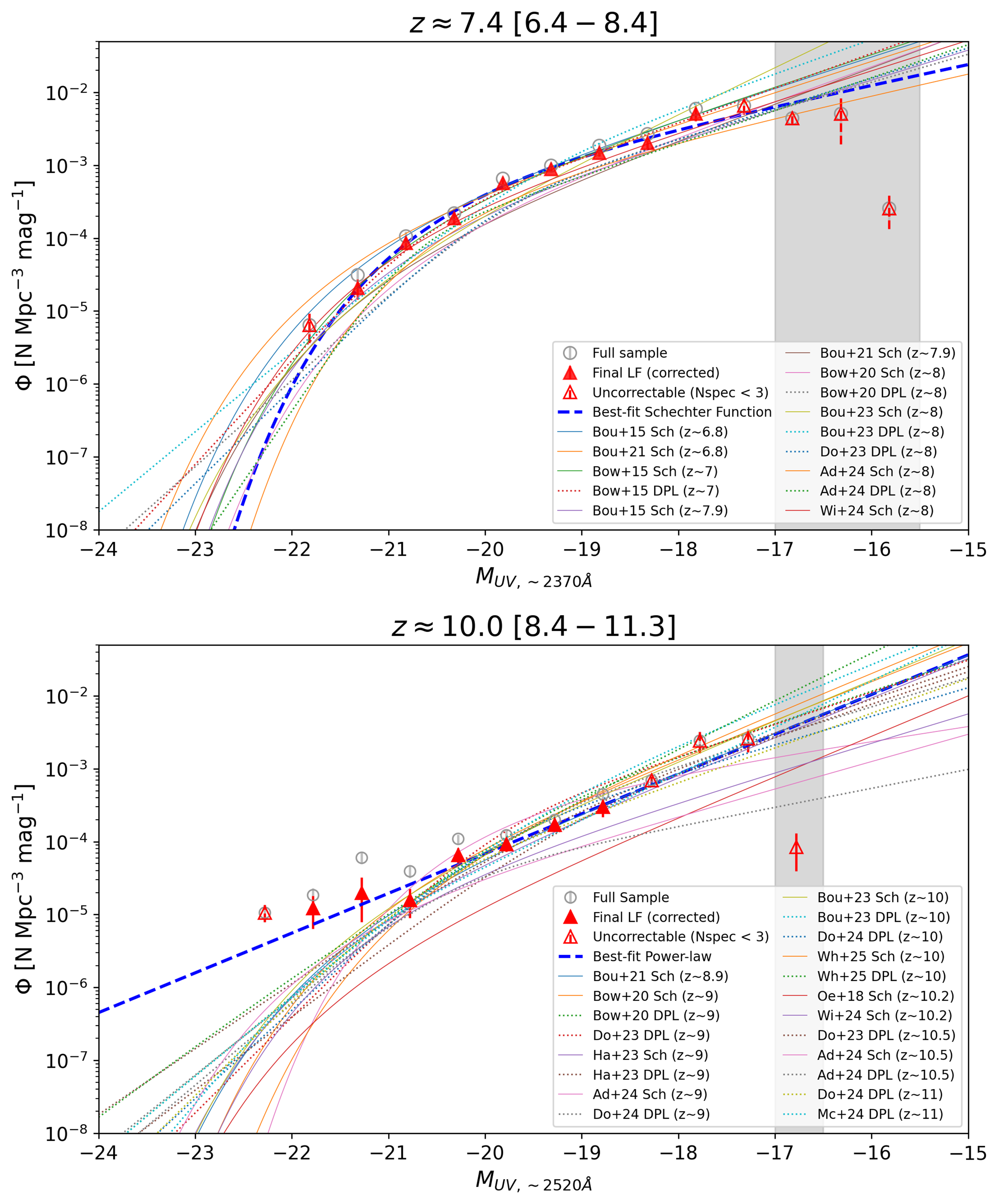}
\caption{Luminosity functions at $z\approx7.4$ (top) and $z\approx10$ (bottom)
derived based on the F090W and the F115W dropouts, respectively. The filled
red triangles indicate $\Phi_{\rm corr}$, and the best-fit analytic form to
these values is shown as a blue dashed curve/line: a Schechter function for 
$z\approx7.4$ and a power law for $z\approx10$. For comparison, the grey open 
circles show $\Phi_{\rm raw}$. The open red triangles are $\Phi_{\rm raw}$ in
the magnitude bins with $N_{\rm spec}<3$ that cannot be corrected for
contamination (see Section~\ref{sec:cont_rate} and Figure~\ref{fig:cont} for 
details) and are excluded from the analytic fit. The thin solid curves show the 
luminosity functions from the literature at comparable redshifts (see
Table~\ref{tab:lf_compare}).
}
\label{fig:lf}
\end{figure*}

The stepwise luminosity functions at $z\approx 7.4$ and $z\approx 10$ thus 
derived are shown in Figure~\ref{fig:lf}, where the filled red triangles 
indicate $\Phi_{\rm corr}$.  
For comparison, the grey open circles
indicating $\Phi_{\rm raw}$ are also shown. The open red triangles are 
$\Phi_{\rm raw}$ in the magnitude bins with less than three spectroscopic 
matches, for which the contamination correction cannot be made.


\begin{table}
\centering
\scriptsize
\begin{tabular}{ccccc}
\toprule
$z$ &
$M_{\rm UV}$ &
$\Phi_{\rm raw}$ &
$\Phi_{\rm corr}$ &
$f_{\rm ctm}$ \\
&
&
\multicolumn{2}{c}{$(10^{-5}\ {\rm Mpc}^{-3}\,{\rm mag}^{-1})$}
& (\%) \\
\midrule
7.40 & -21.82 & $0.65 \pm 0.28$       & ...                         & ... \\
7.40 & -21.32 & $3.13 \pm 0.63$       & $2.09 \pm 0.65$           & $33.3\pm15.7$ \\
7.40 & -20.82 & $10.65 \pm 1.72$      & $8.60 \pm 1.62$          & $19.2\pm7.7$ \\
7.40 & -20.32 & $22.23 \pm 3.64$      & $18.92 \pm 3.31$          & $14.9\pm5.2$ \\
7.40 & -19.82 & $65.83 \pm 4.70$     & $57.34 \pm 4.96$         & $12.9\pm4.3$ \\
7.40 & -19.32 & $100.39 \pm 9.16$    & $89.37 \pm 8.86$        & $11.0\pm3.5$ \\
7.40 & -18.82 & $187.68 \pm 15.15$    & $149.45 \pm 15.86$        & $20.4\pm5.5$ \\
7.40 & -18.32 & $269.51 \pm 34.82$    & $203.95 \pm 32.49$        & $24.3\pm7.1$ \\
7.40 & -17.82 & $597.82 \pm 108.16$  & $512.41 \pm 108.26$      & $14.3\pm9.4$ \\
7.40 & -17.32 & $665.71 \pm 167.17$  & ...                         & ... \\
7.40 & -16.82 & $445.13 \pm 85.98$   & ...                         & ... \\
7.40 & -16.32 & $511.21 \pm 316.20$   & ...                         & ... \\
7.40 & -15.82 & $26.00 \pm 12.59$   & ...                         & ... \\
\midrule
10.00 & -22.28 & $1.07 \pm 0.28$      & ...                         & ... \\
10.00 & -21.78 & $1.85 \pm 0.46$      & $1.23 \pm 0.59$           & $33.3\pm27.2$ \\
10.00 & -21.28 & $6.01 \pm 1.07$     & $2.00 \pm 1.21$           & $66.7\pm19.2$ \\
10.00 & -20.78 & $3.96 \pm 0.77$      & $1.59 \pm 0.69$           & $60.0\pm15.5$ \\
10.00 & -20.28 & $10.97 \pm 1.54$     & $6.58 \pm 1.52$          & $40.0\pm11.0$ \\
10.00 & -19.78 & $12.39 \pm 2.01$     & $9.29 \pm 2.02$          & $25.0\pm10.8$ \\
10.00 & -19.28 & $19.91 \pm 2.85$     & $17.07 \pm 2.78$          & $14.3\pm6.6$ \\
10.00 & -18.78 & $45.20 \pm 6.78$    & $30.13 \pm 8.42$         & $33.3\pm15.7$ \\
10.00 & -18.28 & $69.52 \pm 10.09$   & ...                         & ... \\
10.00 & -17.78 & $244.32 \pm 79.18$  & ...                         & ... \\
10.00 & -17.28 & $261.34 \pm 92.29$  & ...                         & ... \\
10.00 & -16.78 & $4.26 \pm 2.29$  & ...                         & ... \\
\bottomrule
\end{tabular}
\caption{Stepwise luminosity functions at $z\approx 7.4$ and $10$, 
before ($\Phi_{\rm raw}$) and after ($\Phi_{\rm corr}$) correcting for the 
contamination fraction $f_{\rm ctm}$ in each magnitude bin. 
The correction is only applied to the magnitude bins with at least three 
spectroscopically identified objects ($N_{\rm spec}\geq3$). 
For the bins with $N_{\rm spec}<3$, only the uncorrected $\Phi$ values are 
reported. 
}
\label{tab:lf}
\end{table}

\subsection{Analytic Fits}\label{sec:lf_behavior}

Within the luminosity range where our results are well constrained by the 
spectroscopic sample (i.e., the filled red triangles in Figure~\ref{fig:lf}), 
we fit analytic functions to better describe their behaviors. 

At $z\approx7.4$, the contamination-corrected number densities 
$\Phi_{\rm corr}$ show a clear exponential down-turn at the bright end, and the 
data points are well described by a conventional Schechter function,
\begin{equation}
\Phi(M)\,{\rm d}M = (0.4\ln10)\,\Phi_*\,10^{0.4(\alpha+1)(M_*-M)}
\exp\!\left[-10^{0.4(M_*-M)}\right]{\rm d}M
\end{equation}
with the best-fit parameters of $\log_{10}(\Phi_*)=-3.02\pm0.16$, 
$M_*=-20.12\pm0.22$, and $\alpha=-1.70\pm0.16$. This is shown as the dashed
blue curve in the top panel of Figure~\ref{fig:lf}, where the luminosity
functions at $z\approx 7$ and 8 from the literature are shown for comparison
(see also Table~\ref{tab:lf_compare}). Over $M_{\rm UV}\approx -21$ to 
$-18$~mag, ours is broadly consistent with the two pre-JWST $z\approx 7$ 
results of \citet[][]{Bouwens2015, Bouwens2021} but are significantly higher 
than the previous estimates at $z\approx 8$, including those using the JWST
data. This could be broadly explained by the differences in the redshift ranges 
that these individual studies probed as compared to ours. On the other hand, 
our faint-end slope is significantly shallower than any previous estimates at 
$z\approx 7$--8. A decisive derivation of this slope would need reliable 
measurements at $M_{\rm UV}\gtrsim -17.5$~mag, which are still not allowed by 
the existing data: the four bins probing this regime all suffer from very 
few spectroscopic identifications and cannot be used, among which the three 
faintest bins (all based on the magnified dropouts in the lensing cluster 
fields) also suffer from very unreliable incompleteness assessments. 

Our luminosity function at $z\approx10$ behaves very differently from the
$z\approx 7.4$ one. The contamination-corrected number densities in $\log\Phi$
closely follow a straight line across the full luminosity range, 
with no evidence for a down-turn like at $z \approx 7.4$. 
Therefore, we fit using a single power law by removing the exponential 
term from the Schechter function: 
\begin{equation}
\Phi(M)\,{\rm d}M = (0.4\ln10)\,\Phi_*\,10^{0.4(\beta+1)(M_*-M)}\,{\rm d}M
\end{equation}
This form is specifically chosen for ease of comparison of the
power-law slope described by $\beta$ in this equation and the faint-end
slope $\alpha$ in the Schechter function. The choice of $M_*$ is arbitrary in
this case, which only affects the fiducial normalization, i.e., the value of 
$\log_{10}\Phi_*$, but not the power-law slope $\beta$. We adopt 
$M_*=-21.0$~mag, and obtain the best-fit parameters of 
$\log_{10}\Phi_*=-4.67\pm0.08$ and $\beta=-2.36\pm0.16$. 
Such a power-law behavior is mainly driven by the results in the two brightest,
``correctable'' bins ($-21.78$ and $-21.28$~mag), where the number of
spectroscopic identifications is still small ($N_{\rm spec}=3$ and 6,
respectively). Nevertheless, we believe that our results are robust for three 
reasons. First, the bright-end is not dominated by a single field: the 
$M_{\rm UV}=-21.78$ and $-21.28$ bins contain 83 and 123 candidates selected 
from seven fields, respectively. Second, four galaxies in these bins have been 
confirmed: two sources in the Abell 2744 field at $z_{\rm spec}=9.32$ and 
$8.52$ in the $-21.78$~mag bin, and one GOODS-N and one COSMOS source at 
$z_{\rm spec}=10.61$ and $9.81$ in the $-21.28$~mag bin. This shows that 
genuinely luminous high-redshift galaxies are present in both bins. 
Third, the bright-end behavior is not likely caused by an
underestimate of the contamination rate: even if we lower $\Phi_{\rm corr}$ in 
these two bins to the lower bounds allowed by the 1~$\sigma$ errors in the 
contamination rate estimates, the luminosity function is still more consistent
with a power law than a Schechter function. We note that such a power-law
behavior is very similar to that of the $z\approx 12.7$ and $z\approx 17.4$ 
luminosity functions reported in \citet[][]{Yan_highz_23}.


\begin{table*}
    \centering
    \begin{tabular}{lcccccc}
       Literature & Profile & Redshift & $\log_{10}(\Phi_*)$ & $M_*$ & $\alpha$ & $\beta$ \\ \hline
       Bou+15 & Sch & 6.8 & $-3.54$ & $-20.87$ & $-2.06$ & - \\
       Bou+21 & Sch & 6.8 & $-3.72$ & $-21.15$ & $-2.06$ & - \\
       Bow+15 & Sch & 7 & $-3.43$ & $-20.56$ & $-2.09$ & - \\ 
              & DPL & 7 & $-3.66$ & $-20.61$ & $-2.19$ & $-4.60$ \\ 
       Bou+15 & Sch & 7.9 & $-3.68$ & $-20.63$ & $-2.02$ & - \\
       Bou+21 & Sch & 7.9 & $-4.05$ & $-20.93$ & $-2.23$ & - \\
       Bow+20 & Sch & 8 & $-3.72$ & $-20.48$ & $-2.18$ & - \\ 
              & DPL & 8 & $-3.32$ & $-19.80$ & $-1.96$ & $-3.98$ \\
       Bou+23 & Sch & 8 & $-4.24$ & $-21.15$ (fixed) & $-2.58$ & - \\
             & DPL & 8 & $-2.99$ & $-19.67$ (fixed) & $-2.17$ (fixed) & $-3.75$ (fixed) \\ 
        Do+23 & DPL & 8 & $-3.48$ & $-20.02$ & $-2.04$ & $-4.26$ \\ 
       Ad+24 & Sch & 8 & $-3.19$ & $-20.00$ & $-1.74$ & - \\ 
             & DPL & 8 & $-3.95$ & $-20.76$ & $-2.13$ & $-5.87$ \\ 
       Wi+24 & Sch & 8 & $-3.64$ & $-20.75$ & $-2.04$ & - \\ 
       {\bf SY+26} & Sch & 7.4 & $-3.02$ & $-20.12$ & $-1.70$ & - \\ 
        \hline 
       Bou+21 & Sch & 8.9 & $-4.68$ & $-21.15$ (fixed) & $-2.33$ & - \\
       Bow+20 & Sch & 9 & $-4.28$ & $-20.80$ & $-2.31$ & - \\ 
              & DPL & 9 & $-3.55$ & $-19.67$ & $-2.10$ (fixed) & $-3.75$ \\
       Do+23 & DPL & 9 & $-3.68$ & $-19.93$ & $-2.10$ (fixed) & $-4.29$ \\
       Ha+23 & Sch & 9 & $-4.82$ & $-21.24$ & $-2.35$ (fixed) & - \\
             & DPL & 9 & $-3.50$ & $-19.33$ & $-2.10$ (fixed) & $-3.27$ \\ 
       Ad+24 & Sch & 9 & $-3.34$ & $-19.80$ & $-1.50$ & - \\
       Do+24 & DPL & 9 & $-3.63$ & $-19.70$ & $-2.00$ & $-3.81$ \\
       Bou+23 & Sch & 10 & $-4.59$ & $-21.15$ (fixed) & $-2.38$ (fixed) & - \\ 
              & DPL & 10 & $-3.55$ & $-19.67$ (fixed) & $-2.35$ (fixed) & $-3.75$ (fixed) \\ 
       Do+24 & DPL & 10 & $-3.84$ & $-19.98$ & $-1.98$ & $-4.05$ (fixed) \\ 
       Wh+25 & Sch & 10 & $-4.00$ & $-20.32$ & $-2.36$ & - \\
             & DPL & 10 & $-4.31$ & $-20.54$ & $-2.60$ & $-3.49$ \\ 
       Oe+18 & Sch & 10.2 & $-5.38$ & $-21.19$ (fixed) & $-2.38$ & - \\
       Wi+24 & Sch & 10.2 & $-4.53$ & $-20.80$ (fixed) & $-2.0$ (fixed) & - \\ 
       Do+23 & DPL & 10.5 & $-3.48$ & $-19.12$ & $-2.10$ (fixed) & $-3.53$ \\
       Ad+24 & Sch & 10.5 & $-4.83$ & $-21.29$ & $-1.93$ & - \\ 
              & DPL & 10.5 & $-4.41$ & $-20.39$ & $-1.65$ & $-3.67$ \\ 
       Do+24 & DPL & 11 & $-4.49$ & $-20.73$ & $-2.19$ & $-4.29$ \\
       Mc+24 & DPL & 11 & $-5.37$ & $-21.62$ & $-2.59$ & $-4.99$ \\
       {\bf SY+26} & PL & 10 & $-4.67$ & $-21.0$ (chosen) & - & $-2.36$ \\ 
    \hline
    \end{tabular}
    \caption{
    Analytic fits to luminosity functions from this work and previous studies. 
    The works from the literature used in the comparison include 
    Bou+15 \citep[][]{Bouwens2015}, 
    Bou+21 \citep[][]{Bouwens2021}, 
    Bow+15 \citep[][]{Bowler2015}, 
    Bow+20 \citep[][]{Bowler20}, 
    Bou+23 \citep[][]{Bouwens2023}, 
    Do+23 \citep[][]{Donnan23}, 
    Ad+24 \citep[][]{Adams23}, 
    Wi+24 \citep[][]{Willott24}, 
    Ha+23 \citep[][]{Harikane23_uvlf}, 
    Do+24 \citep[][]{Donnan24}, 
    Mc+24 \citep[][]{McLeod2024}, 
    Wh+25 \citep[][]{Whitler2025}, and 
    Oe+18 \citep[][]{Oesch18}. 
    For Ad+24, we show here the values derived using the LM method, which are labeled as ``Sch/LM" 
    and ``DPL/LM" in their Table~4; at $z=10.5$, we take their fitting results when all parameters are freed. 
    }
    \label{tab:lf_compare}
\end{table*}


\subsection{Number Density Evolution from $\mathbf{\MakeLowercase{z}\approx 10}$ to 7.4}\label{sec:discussion}

\begin{figure*}
\centering
\includegraphics[width=0.9\textwidth]{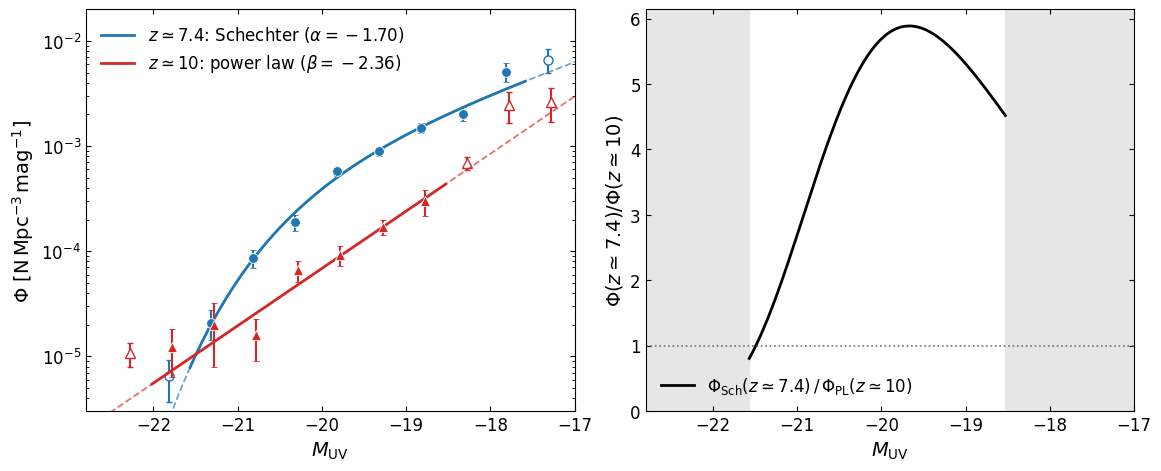}
\caption{Evolution of the galaxy UV luminosity function between $z\approx10$ and $z\approx7.4$. 
{\it Left:} contamination-correction luminosity functions at $z\approx7.4$ (blue circles) and $z\approx10$ (red triangles) 
with their best-fit Schechter and power-law profiles, respectively. 
Open symbols show magnitude bins for which the contamination correction is unreliable due to insufficient numbers of spectroscopically identified sources. 
Solid curves indicate the $M_{\rm UV}$ ranges from the contamination-corrected data points, 
while dashed lines show the extrapolations into brighter and fainter $M_{\rm UV}$ ranges. 
{\it Right:} Ratio of the best-fit Schechter function at $z\approx7.4$ and the best-fit power law at $z\approx10$; 
the light grey regions are the magnitude bins that lie outside of the range constrained at both redshifts. 
The horizontal dotted line shows no evolution, i.e., $\Phi_{z\approx7.4}/\Phi_{z\approx10}=1$. 
}
\label{fig:lf_evo}
\end{figure*}

Figure~\ref{fig:lf_evo} compares the luminosity functions at $z\approx 10$ and
7.4 in the left panel and shows the ratio of the two (using their best-fit 
analytic forms) in the right panel. Apparently, the evolution between these
two redshifts is only minimal at the very bright end 
($M_{\rm UV}\lesssim -21.5$~mag) but quickly becomes substantial at the fainter 
levels. The ratio reaches $\sim6$ near $M_{\rm UV}\approx-19.5$ but then drops 
to $\sim4$--5 at $M_{\rm UV}\approx -18.5$~mag. If the trends of both 
luminosity functions keep the same beyond the luminosity range where we 
currently have reliable constraints, the crossover of the two will happen at 
$M_{\rm UV}\approx -15.72$~mag, fainter than which there would be more
galaxies at $z\approx 10$ than at $z\approx 7.4$. Qualitatively, this could be
understood by hierarchical merging over $\sim$240~Myrs from $z\approx 10$ to
7.4, i.e., more numerous, faint galaxies at $z\approx 10$ merge to become
brighter ones at $z\approx 7.4$. However, such a picture would need to be 
substantiated by observations at $\sim$32~mag in the future.

\section{Summary}\label{sec:summary}

In this work, we present a systematic search for candidate galaxies at 
$6.4\leq z< 11.3$ in nine JWST NIRCam fields totaling $\sim$1,045~arcmin$^2$, 
including seven blank fields and two galaxy cluster fields. Using the dropout 
method, we selected 3,733 F090W dropouts (candidates at $6.4\leq z< 8.4$,
or $z\approx 7.4$) and 2,396 F115W dropouts (candidates at 
$8.4\leq z< 11.3$, or $z\approx 10$), which constitute the largest samples 
of galaxies at these redshifts from JWST after four years of operation. The 
large number of JWST NIRSpec observations available to date have securely 
identified 338 F090W dropouts and 99 F115W dropouts, which afforded us an 
unprecedented spectroscopic subsample to derive magnitude-dependent 
contamination rates. 

Based on the contamination-corrected samples, we derived the galaxy luminosity 
functions at $z\approx7.4$ and $z\approx10$. Over $M_{\rm UV}\approx -21$ to 
$-18$~mag, we find that ours at $z\approx 7.4$ is close to the ones from the
previous studies at $z\approx 7$ and is significantly higher than those at 
$z\approx 8$. This luminosity function can be well fitted by a conventional 
Schechter function, with the best-fit parameters of 
$\log_{10}(\Phi_*)=-3.02\pm0.16$, $M_*=-20.12\pm0.22$, and
$\alpha=-1.70\pm0.16$. This faint-end slope is significantly shallower than
those from any previous studies at $z\approx 7$--8, most of which have
$\alpha \approx -2$. Verifying this result will need more observations 
(both imaging and spectroscopy) at $\sim$30~mag.

Our luminosity function at $z\approx 10$, on the other hand, is consistent with
a power law but not a Schechter function. Using the power-law form of
$\Phi(M)\propto 10^{-0.4(\beta+1)M}$, the best-fit slope is 
$\beta=-2.36\pm0.16$. 
The power-law behavior is mainly driven
by the high number densities at $M_{\rm UV}\approx -21.28$ and $-21.78$~mag; 
given the spectroscopic confirmations in these two bins, we believe that our 
result is reliable. More spectroscopic confirmations at the bright end
($\lesssim 26$~mag) will lend further support to this conclusion.

Comparing these two luminosity functions, we find little evolution over 
$z\approx 10$ to 7.4 at $M_{\rm UV}\lesssim -21.5$~mag but a strong evolution
at $\gtrsim -21$~mag, where the galaxy number densities at $z\approx 7.4$ are 
much higher (up to $\sim$6$\times$) than the ones at $z\approx 10$. However,
due to the much steeper $\beta=-2.36$ at $z\approx 10$ than $\alpha=-1.70$
at $z\approx 7.4$, the number density at $z\approx 10$ would overtake that at
$z\approx 7.4$ at $M\approx -15.72$~mag if both luminosity functions are still
valid in this luminosity regime. Verifying such a picture will need observations
reaching $\sim$32~mag in blank fields.


\section*{Acknowledgements}

We acknowledge the support from the University of Missouri Research Council grant URC-23-029, NASA grant 80NSSC23K0491, and NSF grant AST-2307447. 

This project is based on the observations made with the NASA/ESA/CSA James Webb Space Telescope 
and obtained from the Mikulski Archive for Space Telescopes 
at the Space Telescope Science Institute (STScI), 
which is operated by the Association of Universities for Research in Astronomy, Inc., 
under NASA contract NAS 5-03127 for JWST.

\section*{Data Availability}

All the JWST data used in this paper can be found in MAST at doi:\href{10.17909/mn3r-b684}{10.17909/mn3r-b684}.



\bibliographystyle{mnras}
\bibliography{reference} 




\appendix

\section{Spectroscopically Confirmed Redshifts}\label{sec:all_spec_z}

The individual objects in our spectroscopic sample are summarized in the tables
below. The objects are arranged by their sequential ID (SID) in our dropout
sample. Their ICRS positions (R.A. and Decl. in J2000.0) and F200W magnitudes
($m_{200}$) are based on our measurements, while their redshifts
($z_{\rm spec}$) are from the DJA. Only secure redshifts from the DJA (ranks of
2 or 3) are included in our spectroscopic sample.

\onecolumn
\begin{landscape}

\setlength{\tabcolsep}{3pt}
\small
\begin{longtable}{lcccc@{\hspace{0.18cm}}|@{\hspace{0.18cm}}lcccc@{\hspace{0.18cm}}|@{\hspace{0.18cm}}lcccc}
\caption{F090W-dropouts confirmed at high-$z$. }
\label{tab:f090d_specmatch_highz}\\

\toprule
SID & R.A. & Decl. & $z_{\rm spec}$ & $m_{200}$ &
SID & R.A. & Decl. & $z_{\rm spec}$ & $m_{200}$ &
SID & R.A. & Decl. & $z_{\rm spec}$ & $m_{200}$ \\
\midrule
\endfirsthead
\multicolumn{15}{c}{{\tablename\ \thetable{} -- continued from previous page}} \\
\toprule
SID & R.A. & Decl. & $z_{\rm spec}$ & $m_{200}$ &
SID & R.A. & Decl. & $z_{\rm spec}$ & $m_{200}$ &
SID & R.A. & Decl. & $z_{\rm spec}$ & $m_{200}$ \\
\midrule
\endhead
\midrule
\multicolumn{15}{r}{{Continued on next page}} \\
\endfoot
\bottomrule
\endlastfoot
f090d\_aco2744\_0004 & 3.592421 & -30.432829 & 6.75 & 28.56 & f090d\_aco2744\_0082 & 3.608514 & -30.418542 & 7.29 & 25.26 & f090d\_aco2744\_0090 & 3.593788 & -30.415424 & 6.59 & 26.64 \\
f090d\_aco2744\_0150 & 3.636963 & -30.406361 & 8.52 & 25.57 & f090d\_aco2744\_0151 & 3.608054 & -30.406276 & 6.59 & 28.50 & f090d\_aco2744\_0194 & 3.582350 & -30.402740 & 6.88 & 28.35 \\
f090d\_aco2744\_0206 & 3.579831 & -30.401570 & 7.04 & 28.09 & f090d\_aco2744\_0207 & 3.567765 & -30.401258 & 6.57 & 28.22 & f090d\_aco2744\_0244 & 3.582955 & -30.395234 & 6.87 & 27.98 \\
f090d\_aco2744\_0312 & 3.596080 & -30.385818 & 7.88 & 26.85 & f090d\_aco2744\_0353 & 3.603364 & -30.382237 & 7.87 & 26.02 & f090d\_aco2744\_0356 & 3.603842 & -30.382247 & 7.87 & 26.19 \\
f090d\_aco2744\_0364 & 3.576518 & -30.381087 & 7.09 & 29.36 & f090d\_aco2744\_0368 & 3.606455 & -30.380977 & 7.88 & 27.06 & f090d\_aco2744\_0373 & 3.605245 & -30.380585 & 7.88 & 28.51 \\
f090d\_aco2744\_0375 & 3.606562 & -30.380908 & 7.87 & 26.85 & f090d\_aco2744\_0377 & 3.604516 & -30.380443 & 7.88 & 26.12 & f090d\_aco2744\_0397 & 3.601341 & -30.379205 & 7.89 & 26.44 \\
f090d\_aco2744\_0400 & 3.588974 & -30.378646 & 7.88 & 27.70 & f090d\_aco2744\_0403 & 3.578071 & -30.378655 & 6.83 & 27.66 & f090d\_aco2744\_0425 & 3.587891 & -30.376299 & 7.87 & 28.11 \\
f090d\_aco2744\_0426 & 3.587806 & -30.376278 & 7.87 & 26.95 & f090d\_aco2744\_0472 & 3.569594 & -30.373223 & 7.04 & 29.06 & f090d\_aco2744\_0673 & 3.552065 & -30.360153 & 8.60 & 28.12 \\
f090d\_aco2744\_0709 & 3.605254 & -30.357940 & 7.59 & 28.10 & f090d\_aco2744\_0776 & 3.489865 & -30.354457 & 7.30 & 29.14 & f090d\_aco2744\_0895 & 3.575934 & -30.348039 & 6.72 & 27.62 \\
f090d\_aco2744\_1097 & 3.584738 & -30.338974 & 7.57 & 27.66 & f090d\_aco2744\_1256 & 3.451966 & -30.326594 & 7.43 & 27.88 & f090d\_aco2744\_1358 & 3.470521 & -30.320120 & 7.98 & 27.79 \\
f090d\_aco2744\_1449 & 3.512449 & -30.313281 & 6.50 & 29.28 & f090d\_aco2744\_1492 & 3.469685 & -30.309030 & 8.41 & 28.12 & f090d\_aco2744\_1560 & 3.603287 & -30.382247 & 7.87 & 27.99 \\
f090d\_aco370\_0207 & 39.992144 & -1.570872 & 7.85 & 27.20 & f090d\_aco370\_0217 & 39.982582 & -1.566661 & 7.76 & 28.98 & f090d\_cosmos\_0113 & 150.111202 & 2.199053 & 7.13 & 27.12 \\
f090d\_cosmos\_0155 & 150.103331 & 2.207966 & 6.69 & 26.42 & f090d\_cosmos\_0266 & 150.145118 & 2.226211 & 6.74 & 27.33 & f090d\_cosmos\_0512 & 150.141815 & 2.269258 & 6.85 & 27.29 \\
f090d\_cosmos\_0578 & 150.106484 & 2.278046 & 6.86 & 26.98 & f090d\_cosmos\_0740 & 150.178882 & 2.300010 & 7.16 & 27.87 & f090d\_cosmos\_0779 & 150.165661 & 2.304921 & 6.68 & 27.57 \\
f090d\_cosmos\_1317 & 150.175089 & 2.391361 & 7.95 & 27.61 & f090d\_cosmos\_1454 & 150.115778 & 2.426591 & 7.06 & 27.46 & f090d\_egs\_0006 & 215.057315 & 52.891629 & 7.29 & 27.40 \\
f090d\_egs\_0017 & 215.137762 & 52.949776 & 7.84 & 27.15 & f090d\_egs\_0054 & 215.024946 & 52.874673 & 6.67 & 26.98 & f090d\_egs\_0066 & 215.129888 & 52.949955 & 7.43 & 27.22 \\
f090d\_egs\_0078 & 215.132762 & 52.953248 & 6.93 & 27.90 & f090d\_egs\_0080 & 214.843099 & 52.747893 & 7.51 & 27.37 & f090d\_egs\_0124 & 215.128855 & 52.955193 & 7.78 & 27.63 \\
f090d\_egs\_0155 & 215.037085 & 52.891920 & 7.20 & 26.64 & f090d\_egs\_0216 & 214.890722 & 52.795831 & 6.68 & 28.17 & f090d\_egs\_0218 & 214.880466 & 52.789011 & 7.43 & 28.87 \\
f090d\_egs\_0229 & 215.025532 & 52.892895 & 6.96 & 27.74 & f090d\_egs\_0256 & 214.948821 & 52.842103 & 6.74 & 27.79 & f090d\_egs\_0259 & 214.936598 & 52.833777 & 7.03 & 28.72 \\
f090d\_egs\_0274 & 215.037230 & 52.906727 & 7.19 & 26.95 & f090d\_egs\_0275 & 214.789821 & 52.730808 & 6.73 & 27.32 & f090d\_egs\_0277 & 214.812054 & 52.746760 & 7.47 & 27.13 \\
f090d\_egs\_0281 & 214.853008 & 52.776498 & 6.95 & 28.07 & f090d\_egs\_0290 & 214.813809 & 52.749567 & 6.74 & 28.30 & f090d\_egs\_0302 & 214.944015 & 52.844205 & 8.43 & 27.82 \\
f090d\_egs\_0305 & 214.943876 & 52.844247 & 8.43 & 27.54 & f090d\_egs\_0343 & 214.943565 & 52.845691 & 6.96 & 27.24 & f090d\_egs\_0356 & 214.960287 & 52.858848 & 6.72 & 27.64 \\
f090d\_egs\_0397 & 215.074644 & 52.944736 & 6.93 & 27.65 & f090d\_egs\_0399 & 215.074615 & 52.944664 & 6.93 & 26.73 & f090d\_egs\_0404 & 214.978424 & 52.877414 & 7.01 & 27.34 \\
f090d\_egs\_0424 & 215.126732 & 52.983854 & 6.68 & 26.91 & f090d\_egs\_0432 & 215.077880 & 52.950134 & 7.39 & 27.41 & f090d\_egs\_0440 & 215.081744 & 52.954155 & 6.94 & 27.08 \\
f090d\_egs\_0466 & 215.119658 & 52.982835 & 6.94 & 26.87 & f090d\_egs\_0473 & 214.961112 & 52.871648 & 7.29 & 28.88 & f090d\_egs\_0483 & 214.988554 & 52.891824 & 7.09 & 28.60 \\
f090d\_egs\_0504 & 214.927855 & 52.850021 & 7.48 & 28.09 & f090d\_egs\_0519 & 214.826925 & 52.780450 & 6.93 & 27.66 & f090d\_egs\_0548 & 214.882256 & 52.824668 & 6.77 & 26.99 \\
f090d\_egs\_0699 & 214.943454 & 52.882495 & 7.16 & 28.56 & f090d\_egs\_0708 & 214.883027 & 52.840435 & 7.82 & 26.73 & f090d\_egs\_0728 & 214.961085 & 52.897125 & 7.72 & 28.31 \\
f090d\_egs\_0731 & 214.878444 & 52.838482 & 6.96 & 27.49 & f090d\_egs\_0741 & 214.989054 & 52.919652 & 6.68 & 26.42 & f090d\_egs\_0743 & 214.834018 & 52.810312 & 6.63 & 27.65 \\
f090d\_egs\_0756 & 214.806534 & 52.792772 & 6.99 & 27.48 & f090d\_egs\_0778 & 214.818504 & 52.803544 & 6.73 & 28.71 & f090d\_egs\_0783 & 214.892512 & 52.856909 & 6.73 & 26.85 \\
f090d\_egs\_0790 & 214.963919 & 52.909697 & 7.17 & 28.83 & f090d\_egs\_0849 & 214.771887 & 52.778191 & 6.54 & 28.11 & f090d\_egs\_0881 & 214.833345 & 52.827604 & 7.05 & 28.48 \\
f090d\_egs\_0885 & 214.825106 & 52.822673 & 7.05 & 28.11 & f090d\_egs\_0907 & 214.893669 & 52.874035 & 7.03 & 27.86 & f090d\_egs\_0912 & 214.893974 & 52.874604 & 7.03 & 27.51 \\
f090d\_egs\_0915 & 214.806243 & 52.812729 & 6.72 & 26.62 & f090d\_egs\_0919 & 214.831629 & 52.831522 & 7.05 & 27.90 & f090d\_egs\_0933 & 214.859196 & 52.853552 & 7.11 & 27.72 \\
f090d\_egs\_0934 & 214.859150 & 52.853659 & 7.11 & 27.29 & f090d\_egs\_0961 & 214.967558 & 52.932960 & 8.72 & 26.30 & f090d\_egs\_0967 & 214.892667 & 52.880635 & 6.99 & 27.84 \\
f090d\_egs\_0970 & 214.892530 & 52.880676 & 6.99 & 28.09 & f090d\_egs\_0974 & 214.800191 & 52.814529 & 6.73 & 27.06 & f090d\_egs\_0977 & 214.942489 & 52.914729 & 6.97 & 25.60 \\
f090d\_egs\_0984 & 214.844820 & 52.847599 & 7.82 & 27.84 & f090d\_egs\_1038 & 214.894357 & 52.887114 & 7.49 & 28.19 & f090d\_egs\_1082 & 214.898039 & 52.892979 & 7.48 & 28.56 \\
f090d\_egs\_1093 & 214.900464 & 52.895536 & 6.97 & 27.50 & f090d\_egs\_1110 & 214.813049 & 52.834256 & 7.18 & 27.02 & f090d\_egs\_1128 & 214.830981 & 52.848309 & 7.56 & 28.22 \\
f090d\_egs\_1135 & 214.896656 & 52.895814 & 6.59 & 28.52 & f090d\_egs\_1139 & 214.985508 & 52.958756 & 6.98 & 27.14 & f090d\_egs\_1140 & 214.978326 & 52.953825 & 6.67 & 26.86 \\
f090d\_egs\_1148 & 214.827810 & 52.850628 & 6.93 & 28.17 & f090d\_egs\_1149 & 214.781065 & 52.817440 & 7.69 & 28.33 & f090d\_egs\_1152 & 214.861601 & 52.876053 & 7.17 & 27.88 \\
f090d\_egs\_1154 & 214.861611 & 52.876182 & 7.17 & 27.92 & f090d\_egs\_1161 & 214.881000 & 52.891232 & 6.48 & 26.82 & f090d\_egs\_1176 & 214.836984 & 52.867021 & 7.16 & 27.75 \\
f090d\_egs\_1177 & 214.879002 & 52.896764 & 6.53 & 27.12 & f090d\_egs\_1179 & 214.842574 & 52.871218 & 7.47 & 27.31 & f090d\_egs\_1181 & 214.776663 & 52.824783 & 6.88 & 28.79 \\
f090d\_egs\_1184 & 214.877914 & 52.897691 & 6.54 & 28.30 & f090d\_egs\_1193 & 214.862838 & 52.889395 & 6.65 & 27.92 & f090d\_egs\_1194 & 214.863059 & 52.889449 & 6.58 & 26.43 \\
f090d\_egs\_1195 & 214.967257 & 52.963635 & 6.99 & 27.53 & f090d\_egs\_1196 & 214.825376 & 52.863080 & 7.18 & 28.23 & f090d\_egs\_1201 & 214.908800 & 52.924172 & 6.82 & 99.00 \\
f090d\_egs\_1211 & 214.950944 & 52.957040 & 6.50 & 27.13 & f090d\_egs\_1214 & 214.934675 & 52.947077 & 6.65 & 27.39 & f090d\_egs\_1216 & 214.906480 & 52.927574 & 6.69 & 27.70 \\
f090d\_egs\_1219 & 215.010942 & 53.003013 & 6.67 & 28.72 & f090d\_egs\_1224 & 214.839322 & 52.882586 & 7.01 & 28.80 & f090d\_egs\_1309 & 215.001128 & 53.011278 & 7.10 & 27.88 \\
f090d\_egs\_1319 & 214.869103 & 52.918906 & 6.53 & 26.09 & f090d\_egs\_1321 & 214.994948 & 53.007928 & 7.45 & 27.65 & f090d\_egs\_1330 & 214.866901 & 52.918645 & 6.96 & 28.11 \\
f090d\_gdn\_0025 & 189.235766 & 62.141201 & 6.73 & 28.12 & f090d\_gdn\_0065 & 189.303066 & 62.153751 & 7.27 & 28.18 & f090d\_gdn\_0069 & 189.263552 & 62.154798 & 6.71 & 27.97 \\
f090d\_gdn\_0114 & 189.210654 & 62.163303 & 6.54 & 28.05 & f090d\_gdn\_0150 & 189.211409 & 62.170307 & 8.38 & 27.55 & f090d\_gdn\_0175 & 189.197993 & 62.177026 & 7.14 & 27.91 \\
f090d\_gdn\_0190 & 189.279503 & 62.179746 & 6.83 & 27.21 & f090d\_gdn\_0194 & 189.185527 & 62.179814 & 6.76 & 27.46 & f090d\_gdn\_0204 & 189.217554 & 62.182762 & 7.14 & 28.05 \\
f090d\_gdn\_0239 & 189.232995 & 62.189113 & 8.63 & 28.10 & f090d\_gdn\_0262 & 189.330690 & 62.194853 & 7.60 & 27.55 & f090d\_gdn\_0291 & 189.231993 & 62.202337 & 6.55 & 26.90 \\
f090d\_gdn\_0314 & 189.159107 & 62.205758 & 7.31 & 26.03 & f090d\_gdn\_0328 & 189.221854 & 62.207365 & 6.55 & 27.84 & f090d\_gdn\_0346 & 189.275256 & 62.212447 & 7.44 & 28.38 \\
f090d\_gdn\_0389 & 189.084128 & 62.222019 & 7.19 & 26.40 & f090d\_gdn\_0394 & 189.073697 & 62.223441 & 7.04 & 26.90 & f090d\_gdn\_0400 & 189.186298 & 62.225402 & 6.99 & 28.36 \\
f090d\_gdn\_0419 & 189.170336 & 62.229500 & 6.73 & 27.95 & f090d\_gdn\_0440 & 189.259295 & 62.235467 & 7.14 & 27.97 & f090d\_gdn\_0445 & 189.258899 & 62.237455 & 7.15 & 28.02 \\
f090d\_gdn\_0455 & 189.109154 & 62.238652 & 7.14 & 28.40 & f090d\_gdn\_0456 & 189.109593 & 62.238771 & 7.13 & 28.72 & f090d\_gdn\_0457 & 189.109428 & 62.238790 & 7.13 & 28.09 \\
f090d\_gdn\_0467 & 189.249833 & 62.241223 & 7.15 & 26.16 & f090d\_gdn\_0470 & 189.203072 & 62.242488 & 7.62 & 27.32 & f090d\_gdn\_0500 & 189.205302 & 62.250777 & 6.98 & 26.74 \\
f090d\_gdn\_0514 & 189.245275 & 62.252528 & 6.82 & 28.02 & f090d\_gdn\_0534 & 189.197730 & 62.256967 & 8.28 & 28.33 & f090d\_gdn\_0539 & 189.162543 & 62.258240 & 6.99 & 28.00 \\
f090d\_gdn\_0573 & 189.177812 & 62.265583 & 6.71 & 27.61 & f090d\_gdn\_0605 & 189.138071 & 62.274426 & 6.75 & 27.60 & f090d\_gdn\_0609 & 189.139149 & 62.275292 & 6.73 & 26.85 \\
f090d\_gdn\_0613 & 189.138513 & 62.275597 & 6.75 & 26.74 & f090d\_gdn\_0623 & 189.121114 & 62.277805 & 7.46 & 27.94 & f090d\_gdn\_0645 & 189.148731 & 62.280915 & 6.57 & 27.68 \\
f090d\_gdn\_0646 & 189.069645 & 62.281011 & 7.03 & 26.90 & f090d\_gdn\_0677 & 189.121223 & 62.286398 & 6.67 & 27.56 & f090d\_gdn\_0678 & 189.155315 & 62.286459 & 6.81 & 26.82 \\
f090d\_gdn\_0683 & 189.183360 & 62.287730 & 7.03 & 27.58 & f090d\_gdn\_0705 & 189.177308 & 62.291062 & 6.73 & 26.82 & f090d\_gdn\_0735 & 189.120324 & 62.294855 & 6.59 & 27.78 \\
f090d\_gdn\_0763 & 189.165064 & 62.300199 & 6.56 & 26.58 & f090d\_gdn\_0767 & 189.328376 & 62.301846 & 7.08 & 28.85 & f090d\_gdn\_0772 & 189.157802 & 62.302335 & 7.51 & 26.87 \\
f090d\_gdn\_0773 & 189.157975 & 62.302388 & 7.49 & 26.35 & f090d\_gdn\_0776 & 189.439658 & 62.302376 & 7.57 & 26.69 & f090d\_gdn\_0781 & 189.188396 & 62.303043 & 6.73 & 26.91 \\
f090d\_gdn\_0855 & 189.166254 & 62.316504 & 6.57 & 27.20 & f090d\_gdn\_0958 & 189.408058 & 62.339677 & 7.11 & 26.72 & f090d\_gds\_0020 & 53.103269 & -27.905716 & 6.57 & 26.63 \\
f090d\_gds\_0032 & 53.119870 & -27.901526 & 7.95 & 29.23 & f090d\_gds\_0090 & 53.103898 & -27.890529 & 7.96 & 26.14 & f090d\_gds\_0101 & 53.081004 & -27.886828 & 7.19 & 29.13 \\
f090d\_gds\_0102 & 53.066640 & -27.886523 & 7.96 & 28.29 & f090d\_gds\_0119 & 53.099384 & -27.880318 & 7.95 & 27.28 & f090d\_gds\_0123 & 53.075766 & -27.879318 & 8.21 & 27.24 \\
f090d\_gds\_0131 & 53.053676 & -27.877817 & 7.89 & 27.81 & f090d\_gds\_0139 & 53.101002 & -27.875753 & 7.47 & 28.80 & f090d\_gds\_0157 & 53.089279 & -27.872631 & 8.23 & 28.11 \\
f090d\_gds\_0198 & 53.143202 & -27.866356 & 7.45 & 28.02 & f090d\_gds\_0202 & 53.060529 & -27.865951 & 7.88 & 27.74 & f090d\_gds\_0217 & 53.060281 & -27.863469 & 7.89 & 27.54 \\
f090d\_gds\_0223 & 53.113740 & -27.862305 & 7.96 & 28.76 & f090d\_gds\_0241 & 53.086451 & -27.859130 & 7.95 & 27.62 & f090d\_gds\_0254 & 53.119966 & -27.856381 & 7.66 & 28.05 \\
f090d\_gds\_0256 & 53.120174 & -27.856267 & 7.65 & 27.60 & f090d\_gds\_0258 & 53.075387 & -27.855135 & 7.22 & 27.90 & f090d\_gds\_0260 & 53.087416 & -27.854582 & 7.95 & 28.74 \\
f090d\_gds\_0293 & 53.106009 & -27.848156 & 6.91 & 26.64 & f090d\_gds\_0303 & 53.086400 & -27.845463 & 7.38 & 28.54 & f090d\_gds\_0354 & 53.136712 & -27.837395 & 8.23 & 27.85 \\
f090d\_gds\_0362 & 53.157083 & -27.837018 & 7.67 & 27.05 & f090d\_gds\_0371 & 53.059194 & -27.835230 & 7.75 & 28.40 & f090d\_gds\_0445 & 53.086274 & -27.823893 & 7.60 & 27.54 \\
f090d\_gds\_0460 & 53.099058 & -27.820441 & 6.89 & 26.81 & f090d\_gds\_0560 & 53.155059 & -27.801703 & 7.34 & 27.97 & f090d\_gds\_0610 & 53.118153 & -27.792940 & 6.79 & 27.79 \\
f090d\_gds\_0612 & 53.177344 & -27.792056 & 6.80 & 27.23 & f090d\_gds\_0614 & 53.135593 & -27.791785 & 7.76 & 27.56 & f090d\_gds\_0626 & 53.164801 & -27.788192 & 7.24 & 27.23 \\
f090d\_gds\_0634 & 53.127141 & -27.786574 & 6.88 & 27.77 & f090d\_gds\_0644 & 53.161682 & -27.785314 & 7.23 & 26.80 & f090d\_gds\_0652 & 53.145524 & -27.783720 & 6.88 & 27.48 \\
f090d\_gds\_0663 & 53.183410 & -27.790895 & 7.43 & 25.83 & f090d\_gds\_0670 & 53.138020 & -27.781790 & 7.14 & 28.62 & f090d\_gds\_0671 & 53.128211 & -27.781439 & 7.04 & 29.43 \\
f090d\_gds\_0685 & 53.186248 & -27.778968 & 7.27 & 27.41 & f090d\_gds\_0686 & 53.169021 & -27.778762 & 6.63 & 28.45 & f090d\_gds\_0718 & 53.165522 & -27.772590 & 7.24 & 29.04 \\
f090d\_gds\_0726 & 53.180097 & -27.771352 & 7.28 & 28.14 & f090d\_gds\_0734 & 53.181460 & -27.769424 & 7.44 & 28.79 & f090d\_gds\_0736 & 53.118297 & -27.768931 & 7.20 & 28.37 \\
f090d\_gds\_0747 & 53.156801 & -27.767076 & 7.98 & 28.89 & f090d\_gds\_0756 & 53.117276 & -27.764008 & 6.93 & 27.47 & f090d\_gds\_0765 & 53.116283 & -27.761875 & 6.79 & 27.67 \\
f090d\_gds\_0773 & 53.133444 & -27.760318 & 7.66 & 26.27 & f090d\_gds\_0791 & 53.169486 & -27.753230 & 6.62 & 27.57 & f090d\_gds\_0812 & 53.150026 & -27.744924 & 6.57 & 26.42 \\
f090d\_gds\_0814 & 53.132827 & -27.744331 & 6.58 & 28.09 & f090d\_gds\_0823 & 53.172563 & -27.743845 & 7.57 & 25.86 & f090d\_gds\_0854 & 53.168665 & -27.737118 & 7.24 & 28.08 \\
f090d\_uds\_0025 & 34.310283 & -5.284374 & 7.31 & 27.67 & f090d\_uds\_0028 & 34.399924 & -5.283962 & 7.35 & 25.80 & f090d\_uds\_0031 & 34.398005 & -5.283214 & 7.29 & 27.75 \\
f090d\_uds\_0074 & 34.439979 & -5.276728 & 7.29 & 27.17 & f090d\_uds\_0084 & 34.351419 & -5.275284 & 7.51 & 27.32 & f090d\_uds\_0095 & 34.359509 & -5.273383 & 7.50 & 27.86 \\
f090d\_uds\_0133 & 34.349673 & -5.268920 & 6.52 & 27.83 & f090d\_uds\_0135 & 34.280502 & -5.268345 & 7.43 & 26.93 & f090d\_uds\_0148 & 34.424056 & -5.266259 & 7.29 & 27.65 \\
f090d\_uds\_0161 & 34.424970 & -5.265197 & 7.28 & 26.91 & f090d\_uds\_0195 & 34.422223 & -5.259280 & 7.29 & 25.80 & f090d\_uds\_0203 & 34.372386 & -5.257023 & 7.36 & 27.23 \\
f090d\_uds\_0227 & 34.310885 & -5.254371 & 6.61 & 27.10 & f090d\_uds\_0234 & 34.388631 & -5.253537 & 6.85 & 28.51 & f090d\_uds\_0264 & 34.280376 & -5.248375 & 6.80 & 26.05 \\
f090d\_uds\_0274 & 34.445410 & -5.244891 & 7.25 & 26.89 & f090d\_uds\_0293 & 34.281588 & -5.237553 & 7.10 & 26.54 & f090d\_uds\_0295 & 34.305922 & -5.236660 & 6.81 & 27.56 \\
f090d\_uds\_0302 & 34.254911 & -5.232819 & 6.64 & 27.26 & f090d\_uds\_0313 & 34.255719 & -5.226793 & 7.27 & 26.35 & f090d\_uds\_0327 & 34.314291 & -5.223235 & 7.12 & 27.29 \\
f090d\_uds\_0330 & 34.320343 & -5.222708 & 8.27 & 28.05 & f090d\_uds\_0352 & 34.309980 & -5.209171 & 7.46 & 27.40 & f090d\_uds\_0358 & 34.279035 & -5.204370 & 6.53 & 26.59 \\
f090d\_uds\_0364 & 34.313698 & -5.203067 & 7.46 & 26.10 & f090d\_uds\_0377 & 34.283608 & -5.195990 & 7.35 & 27.16 & f090d\_uds\_0455 & 34.472148 & -5.168135 & 6.62 & 26.49 \\
f090d\_uds\_0462 & 34.355364 & -5.167166 & 7.75 & 27.03 & f090d\_uds\_0490 & 34.486709 & -5.160084 & 7.79 & 26.42 & f090d\_uds\_0496 & 34.356933 & -5.159034 & 7.36 & 27.04 \\
f090d\_uds\_0527 & 34.262843 & -5.152969 & 6.97 & 27.97 & f090d\_uds\_0540 & 34.479247 & -5.150566 & 6.72 & 27.36 & f090d\_uds\_0577 & 34.489850 & -5.145780 & 6.59 & 26.50 \\
f090d\_uds\_0587 & 34.442781 & -5.143489 & 6.59 & 27.46 & f090d\_uds\_0602 & 34.353071 & -5.139179 & 7.41 & 27.76 & f090d\_uds\_0626 & 34.413885 & -5.134783 & 6.76 & 26.26 \\
f090d\_uds\_0627 & 34.402488 & -5.134459 & 7.92 & 27.34 & f090d\_uds\_0632 & 34.401009 & -5.134249 & 7.76 & 26.82 & f090d\_uds\_0633 & 34.400912 & -5.134247 & 7.76 & 27.45 \\
f090d\_uds\_0639 & 34.425872 & -5.133704 & 6.96 & 27.57 & f090d\_uds\_0656 & 34.267547 & -5.131739 & 8.03 & 26.07 & f090d\_uds\_0666 & 34.425506 & -5.130587 & 7.57 & 27.19 \\
f090d\_uds\_0698 & 34.320501 & -5.125936 & 6.86 & 26.91 & f090d\_uds\_0714 & 34.446662 & -5.123978 & 7.19 & 27.07 & f090d\_uds\_0736 & 34.475114 & -5.119051 & 6.81 & 27.87 \\
f090d\_uds\_0748 & 34.380352 & -5.116093 & 7.21 & 27.61 &  &  &  &  &  &  &  &  &  &  \\
\end{longtable}

\normalsize
\setlength{\tabcolsep}{6pt}

\end{landscape}
\twocolumn


\onecolumn
\begin{landscape}

\setlength{\tabcolsep}{3pt}
\small
\begin{longtable}{lcccc@{\hspace{0.18cm}}|@{\hspace{0.18cm}}lcccc@{\hspace{0.18cm}}|@{\hspace{0.18cm}}lcccc}
\caption{F090W-dropouts identified as interlopers ($z_{\rm spec}< 6.4$). }
\label{tab:f090d_specmatch_lowz}\\

\toprule
SID & R.A. & Decl. & $z_{\rm spec}$ & $m_{200}$ &
SID & R.A. & Decl. & $z_{\rm spec}$ & $m_{200}$ &
SID & R.A. & Decl. & $z_{\rm spec}$ & $m_{200}$ \\
\midrule
\endfirsthead
\multicolumn{15}{c}{{\tablename\ \thetable{} -- continued from previous page}} \\
\toprule
SID & R.A. & Decl. & $z_{\rm spec}$ & $m_{200}$ &
SID & R.A. & Decl. & $z_{\rm spec}$ & $m_{200}$ &
SID & R.A. & Decl. & $z_{\rm spec}$ & $m_{200}$ \\
\midrule
\endhead
\midrule
\multicolumn{15}{r}{{Continued on next page}} \\
\endfoot
\bottomrule
\endlastfoot

f090d\_aco2744\_0024 & 3.586770 & -30.429603 & 6.14 & 28.45 & f090d\_aco2744\_0123 & 3.614090 & -30.410448 & 6.33 & 25.91 & f090d\_aco2744\_0160 & 3.612429 & -30.405545 & 6.33 & 27.23 \\
f090d\_aco2744\_0197 & 3.567407 & -30.402567 & 2.58 & 25.48 & f090d\_aco2744\_0399 & 3.628746 & -30.378908 & 6.35 & 27.83 & f090d\_aco2744\_0465 & 3.605815 & -30.373912 & 6.37 & 27.07 \\
f090d\_aco2744\_0480 & 3.579887 & -30.372896 & 6.38 & 27.54 & f090d\_aco2744\_0502 & 3.604637 & -30.371153 & 6.38 & 26.80 & f090d\_aco2744\_1266 & 3.560641 & -30.326106 & 6.18 & 28.17 \\
f090d\_cosmos\_0247 & 150.070475 & 2.222711 & 1.33 & 26.55 & f090d\_cosmos\_0332 & 150.148404 & 2.239830 & 5.66 & 28.83 & f090d\_cosmos\_0442 & 150.149379 & 2.258397 & 5.51 & 27.67 \\
f090d\_cosmos\_0651 & 150.065153 & 2.287079 & 1.22 & 26.72 & f090d\_cosmos\_0809 & 150.160876 & 2.309140 & 6.08 & 28.26 & f090d\_cosmos\_1265 & 150.094840 & 2.375537 & 1.26 & 28.64 \\
f090d\_egs\_0036 & 214.971326 & 52.835070 & 0.07 & 29.67 & f090d\_egs\_0364 & 215.120052 & 52.972561 & 4.39 & 27.70 & f090d\_egs\_0426 & 214.874590 & 52.805527 & 6.13 & 27.60 \\
f090d\_egs\_0443 & 215.032115 & 52.918975 & 6.18 & 26.18 & f090d\_egs\_0456 & 214.850370 & 52.791516 & 5.68 & 28.24 & f090d\_egs\_0668 & 214.942910 & 52.877765 & 6.16 & 28.86 \\
f090d\_egs\_0684 & 214.984283 & 52.908861 & 6.15 & 28.02 & f090d\_egs\_0715 & 214.804960 & 52.784945 & 6.11 & 27.33 & f090d\_egs\_0723 & 214.850093 & 52.817512 & 6.27 & 26.98 \\
f090d\_egs\_0785 & 214.811606 & 52.799854 & 6.10 & 28.38 & f090d\_egs\_0861 & 214.977729 & 52.927079 & 6.15 & 28.24 & f090d\_egs\_0865 & 214.905570 & 52.876668 & 5.77 & 27.58 \\
f090d\_egs\_0926 & 214.826604 & 52.829948 & 6.07 & 28.26 & f090d\_egs\_0944 & 214.825189 & 52.830378 & 6.10 & 28.91 & f090d\_egs\_0950 & 214.825047 & 52.830494 & 6.10 & 27.17 \\
f090d\_egs\_1115 & 214.864520 & 52.870994 & 5.31 & 26.60 & f090d\_egs\_1137 & 214.795795 & 52.824212 & 5.85 & 29.22 & f090d\_egs\_1147 & 214.797391 & 52.828940 & 1.99 & 28.25 \\
f090d\_egs\_1192 & 214.770612 & 52.823324 & 6.05 & 28.29 & f090d\_egs\_1228 & 214.860050 & 52.898138 & 6.23 & 27.85 & f090d\_gdn\_0421 & 189.396346 & 62.230178 & 6.32 & 27.92 \\
f090d\_gdn\_0667 & 189.176413 & 62.283684 & 4.16 & 28.73 & f090d\_gds\_0040 & 53.079324 & -27.900794 & 6.27 & 25.56 & f090d\_gds\_0062 & 53.120623 & -27.896648 & 4.62 & 28.87 \\
f090d\_gds\_0112 & 53.092745 & -27.882570 & 6.19 & 25.93 & f090d\_gds\_0458 & 53.144097 & -27.820497 & 1.74 & 28.00 & f090d\_gds\_0603 & 53.115404 & -27.795730 & 2.04 & 27.30 \\
f090d\_gds\_0700 & 53.156157 & -27.775679 & 6.10 & 27.05 & f090d\_gds\_0749 & 53.172622 & -27.767008 & 5.98 & 27.17 & f090d\_gds\_0771 & 53.155116 & -27.760657 & 6.31 & 27.24 \\
f090d\_gds\_0775 & 53.217661 & -27.760972 & 1.75 & 25.70 & f090d\_gds\_0819 & 53.170587 & -27.744682 & 4.12 & 26.86 & f090d\_uds\_0054 & 34.273786 & -5.279692 & 6.36 & 26.42 \\
f090d\_uds\_0363 & 34.444845 & -5.203133 & 5.77 & 27.63 & f090d\_uds\_0378 & 34.467117 & -5.194480 & 6.10 & 27.15 & f090d\_uds\_0523 & 34.258299 & -5.153343 & 5.05 & 26.72 \\
f090d\_uds\_0565 & 34.228340 & -5.147435 & 6.20 & 26.23 & f090d\_uds\_0620 & 34.349468 & -5.136212 & 6.08 & 28.01 & f090d\_uds\_0643 & 34.407435 & -5.133219 & 1.47 & 28.71 \\
f090d\_uds\_0655 & 34.269177 & -5.132368 & 6.19 & 26.38 &  &  &  &  &  &  &  &  &  &  \\

\end{longtable}

\normalsize
\setlength{\tabcolsep}{6pt}

\end{landscape}
\twocolumn


\onecolumn
\begin{landscape}

\setlength{\tabcolsep}{3pt}
\small
\begin{longtable}{lcccc@{\hspace{0.18cm}}|@{\hspace{0.18cm}}lcccc@{\hspace{0.18cm}}|@{\hspace{0.18cm}}lcccc}
\caption{F115W-dropouts confirmed at high-$z$. }
\label{tab:f115d_specmatch_highz}\\

\toprule
SID & RA & Dec & $z_{\rm spec}$ & $m_{277}$ &
SID & RA & Dec & $z_{\rm spec}$ & $m_{277}$ &
SID & RA & Dec & $z_{\rm spec}$ & $m_{277}$ \\
\midrule
\endfirsthead
\multicolumn{15}{c}{{\tablename\ \thetable{} -- continued from previous page}} \\
\toprule
SID & RA & Dec & $z_{\rm spec}$ & $m_{277}$ &
SID & RA & Dec & $z_{\rm spec}$ & $m_{277}$ &
SID & RA & Dec & $z_{\rm spec}$ & $m_{277}$ \\
\midrule
\endhead
\midrule
\multicolumn{15}{r}{{Continued on next page}} \\
\endfoot
\bottomrule
\endlastfoot

f115d\_aco2744\_0054 & 3.617195 & -30.425537 & 9.32 & 25.33 & f115d\_aco2744\_0077 & 3.636963 & -30.406361 & 8.52 & 25.49 & f115d\_aco2744\_0094 & 3.592497 & -30.401463 & 9.80 & 27.78 \\
f115d\_aco2744\_0129 & 3.568117 & -30.383054 & 9.57 & 29.08 & f115d\_aco2744\_0145 & 3.567069 & -30.377867 & 10.06 & 27.23 & f115d\_aco2744\_0196 & 3.590108 & -30.359742 & 10.38 & 27.58 \\
f115d\_aco2744\_0225 & 3.507310 & -30.343206 & 10.07 & 28.56 & f115d\_aco2744\_0254 & 3.451364 & -30.320721 & 10.40 & 27.95 & f115d\_aco2744\_0258 & 3.454711 & -30.316892 & 9.52 & 28.35 \\
f115d\_aco2744\_0266 & 3.469685 & -30.309030 & 8.41 & 28.28 & f115d\_cosmos\_0103 & 150.116593 & 2.197096 & 9.81 & 26.25 & f115d\_cosmos\_0632 & 150.065672 & 2.287486 & 8.87 & 28.37 \\
f115d\_cosmos\_0798 & 150.090334 & 2.306899 & 9.30 & 28.40 & f115d\_cosmos\_1316 & 150.159804 & 2.379295 & 8.94 & 27.95 & f115d\_cosmos\_1350 & 150.126400 & 2.383787 & 8.87 & 26.81 \\
f115d\_cosmos\_1671 & 150.160301 & 2.464051 & 8.85 & 28.05 & f115d\_egs\_0009 & 214.996441 & 52.854171 & 8.83 & 28.27 & f115d\_egs\_0041 & 214.811846 & 52.737127 & 9.57 & 27.41 \\
f115d\_egs\_0064 & 214.817117 & 52.748355 & 9.94 & 27.47 & f115d\_egs\_0094 & 214.932124 & 52.841894 & 10.59 & 28.05 & f115d\_egs\_0140 & 214.897263 & 52.843876 & 9.00 & 28.68 \\
f115d\_egs\_0181 & 214.977384 & 52.926504 & 9.04 & 27.33 & f115d\_egs\_0185 & 214.732534 & 52.758087 & 9.85 & 27.99 & f115d\_egs\_0188 & 214.968727 & 52.929651 & 8.72 & 27.30 \\
f115d\_egs\_0194 & 214.938665 & 52.911755 & 8.76 & 27.03 & f115d\_egs\_0196 & 214.967558 & 52.932960 & 8.72 & 26.51 & f115d\_egs\_0223 & 215.011716 & 52.988309 & 8.82 & 27.30 \\
f115d\_egs\_0237 & 214.994410 & 52.989386 & 8.80 & 27.84 & f115d\_egs\_0248 & 214.985119 & 52.990471 & 8.80 & 27.34 & f115d\_egs\_0250 & 214.844796 & 52.892118 & 8.69 & 26.91 \\
f115d\_gdn\_0092 & 189.180518 & 62.180469 & 8.63 & 27.63 & f115d\_gdn\_0139 & 189.271856 & 62.195188 & 8.67 & 28.13 & f115d\_gdn\_0154 & 189.217710 & 62.199500 & 9.74 & 27.89 \\
f115d\_gdn\_0184 & 189.239809 & 62.210840 & 9.63 & 28.21 & f115d\_gdn\_0185 & 189.239233 & 62.210974 & 9.78 & 28.47 & f115d\_gdn\_0186 & 189.239131 & 62.210939 & 9.78 & 28.12 \\
f115d\_gdn\_0198 & 189.171234 & 62.214761 & 8.61 & 27.97 & f115d\_gdn\_0213 & 189.158262 & 62.221363 & 9.07 & 27.78 & f115d\_gdn\_0261 & 189.106062 & 62.242047 & 10.61 & 26.10 \\
f115d\_gdn\_0356 & 189.142207 & 62.284582 & 9.22 & 28.49 & f115d\_gdn\_0371 & 189.138334 & 62.289855 & 9.26 & 28.05 & f115d\_gds\_0080 & 53.101041 & -27.883044 & 8.81 & 28.36 \\
f115d\_gds\_0131 & 53.124939 & -27.868289 & 8.51 & 28.04 & f115d\_gds\_0182 & 53.139145 & -27.848420 & 9.06 & 27.46 & f115d\_gds\_0197 & 53.133586 & -27.844920 & 9.05 & 26.74 \\
f115d\_gds\_0211 & 53.072086 & -27.841872 & 8.70 & 28.44 & f115d\_gds\_0220 & 53.067690 & -27.837804 & 9.06 & 28.40 & f115d\_gds\_0243 & 53.158587 & -27.834010 & 8.84 & 27.87 \\
f115d\_gds\_0358 & 53.176774 & -27.817689 & 8.56 & 28.58 & f115d\_gds\_0360 & 53.176007 & -27.817324 & 8.68 & 28.88 & f115d\_gds\_0362 & 53.168705 & -27.816909 & 8.75 & 30.23 \\
f115d\_gds\_0425 & 53.167331 & -27.807440 & 9.69 & 28.39 & f115d\_gds\_0457 & 53.169107 & -27.802849 & 8.93 & 29.82 & f115d\_gds\_0464 & 53.164446 & -27.802106 & 8.47 & 28.80 \\
f115d\_gds\_0655 & 53.175495 & -27.780564 & 9.71 & 27.90 & f115d\_gds\_0712 & 53.112402 & -27.774540 & 9.43 & 27.46 & f115d\_gds\_0716 & 53.164738 & -27.774552 & 11.20 & 28.19 \\
f115d\_gds\_0733 & 53.158806 & -27.773414 & 10.31 & 29.05 & f115d\_uds\_0227 & 34.330292 & -5.281234 & 8.56 & 27.84 & f115d\_uds\_0301 & 34.434938 & -5.271138 & 9.30 & 27.02 \\
f115d\_uds\_0671 & 34.465138 & -5.217731 & 9.21 & 27.02 & f115d\_uds\_0849 & 34.460261 & -5.185007 & 9.27 & 27.14 & f115d\_uds\_1053 & 34.305813 & -5.154315 & 9.25 & 27.62 \\
f115d\_uds\_1172 & 34.282246 & -5.142846 & 10.12 & 28.18 & f115d\_uds\_1179 & 34.249790 & -5.142097 & 9.26 & 28.02 & f115d\_uds\_1247 & 34.249408 & -5.130819 & 9.26 & 28.02 \\
f115d\_uds\_1442 & 34.436044 & -5.101404 & 9.28 & 28.24 &  &  &  &  &  &  &  &  &  &  \\

\end{longtable}

\normalsize
\setlength{\tabcolsep}{6pt}

\end{landscape}
\twocolumn


\onecolumn
\begin{landscape}

\setlength{\tabcolsep}{3pt}
\small
\begin{longtable}{lcccc@{\hspace{0.18cm}}|@{\hspace{0.18cm}}lcccc@{\hspace{0.18cm}}|@{\hspace{0.18cm}}lcccc}
\caption{F115W-dropouts identified as interlopers ($z_{\rm spec}< 8.4$). }
\label{tab:f115d_specmatch_lowz}\\

\toprule
SID & RA & Dec & $z_{\rm spec}$ & $m_{277}$ &
SID & RA & Dec & $z_{\rm spec}$ & $m_{277}$ &
SID & RA & Dec & $z_{\rm spec}$ & $m_{277}$ \\
\midrule
\endfirsthead
\multicolumn{15}{c}{{\tablename\ \thetable{} -- continued from previous page}} \\
\toprule
SID & RA & Dec & $z_{\rm spec}$ & $m_{277}$ &
SID & RA & Dec & $z_{\rm spec}$ & $m_{277}$ &
SID & RA & Dec & $z_{\rm spec}$ & $m_{277}$ \\
\midrule
\endhead
\midrule
\multicolumn{15}{r}{{Continued on next page}} \\
\endfoot
\bottomrule
\endlastfoot

f115d\_aco370\_0069 & 39.993521 & -1.610022 & 2.33 & 28.34 & f115d\_aco370\_0176 & 39.994979 & -1.574812 & 8.25 & 28.02 & f115d\_cosmos\_0211 & 150.146680 & 2.219945 & 2.15 & 27.77 \\
f115d\_cosmos\_0303 & 150.163238 & 2.236092 & 2.50 & 26.20 & f115d\_cosmos\_0359 & 150.157392 & 2.244933 & 2.51 & 27.10 & f115d\_cosmos\_0398 & 150.073662 & 2.251317 & 2.11 & 28.37 \\
f115d\_cosmos\_0404 & 150.104834 & 2.252825 & 2.01 & 27.07 & f115d\_cosmos\_0508 & 150.149317 & 2.270940 & 2.21 & 28.21 & f115d\_egs\_0047 & 214.989626 & 52.866572 & 8.28 & 26.80 \\
f115d\_egs\_0089 & 215.120052 & 52.972561 & 4.39 & 27.83 & f115d\_egs\_0166 & 214.876482 & 52.844072 & 2.29 & 28.80 & f115d\_gdn\_0055 & 189.211409 & 62.170307 & 8.38 & 27.70 \\
f115d\_gds\_0761 & 53.183521 & -27.770070 & 8.39 & 27.50 & f115d\_gds\_0783 & 53.156440 & -27.767179 & 6.87 & 29.53 & f115d\_nexus\_0593 & 268.471712 & 65.197504 & 2.35 & 26.94 \\
f115d\_nexus\_0611 & 268.367259 & 65.200770 & 2.64 & 26.13 & f115d\_uds\_0108 & 34.322051 & -5.301306 & 2.39 & 26.95 & f115d\_uds\_0147 & 34.348874 & -5.295278 & 6.07 & 26.97 \\
f115d\_uds\_0358 & 34.422097 & -5.263284 & 8.20 & 27.39 & f115d\_uds\_0368 & 34.270223 & -5.262295 & 5.67 & 26.65 & f115d\_uds\_0379 & 34.421337 & -5.260054 & 8.19 & 28.71 \\
f115d\_uds\_0405 & 34.423101 & -5.257343 & 7.28 & 27.04 & f115d\_uds\_0527 & 34.243202 & -5.239762 & 3.20 & 25.58 & f115d\_uds\_0703 & 34.294582 & -5.211257 & 1.31 & 26.17 \\
f115d\_uds\_0807 & 34.460861 & -5.193651 & 1.70 & 27.33 & f115d\_uds\_1022 & 34.444215 & -5.157913 & 6.58 & 27.63 & f115d\_uds\_1061 & 34.436128 & -5.153462 & 3.08 & 25.45 \\
f115d\_uds\_1063 & 34.490024 & -5.152469 & 2.31 & 27.33 & f115d\_uds\_1148 & 34.419291 & -5.145968 & 8.27 & 26.65 & f115d\_uds\_1176 & 34.265372 & -5.142375 & 5.59 & 27.49 \\
f115d\_uds\_1403 & 34.261369 & -5.108351 & 2.02 & 28.26 & f115d\_uds\_1414 & 34.377384 & -5.106737 & 7.75 & 26.31 &  &  &  &  &  \\

\end{longtable}

\normalsize
\setlength{\tabcolsep}{6pt}

\end{landscape}
\twocolumn

\section{Assessing SED-Fitting-Based Sample Purification}\label{append:sed-purify} 

In the pre-JWST era, it was a common practice in dropout selection to use 
colour-colour diagram(s) involving one (or two) passband(s) to the redder side of 
the shift-in band to reject low-$z$ interlopers under the assumption that their 
colours are usually redder than genuine high-$z$ objects. The slew of NIRCam 
bands makes it possible to replace such colour-colour diagnostics by SED fitting, 
i.e., rejecting low-$z$ interlopers by their photometric redshifts
($z_{\rm phot}$). The hidden assumption is that the templates in use are
sufficient to describe the SEDs of high-$z$ galaxies. This SED-fitting-based 
``sample purification'' approach has been adopted in 
\citet[][]{Yan_highz_23}.

In this work, our spectroscopic sample is large enough that we were able to 
assess the contamination rate directly without assuming the colours (or SEDs) of
high-$z$ objects and low-$z$ interlopers. Nonetheless, it is valuable to
test how well the SED-fitting-based purification could perform, as such a method
(or its variant) could be the only approach to derive contamination rate in 
cases where no spectroscopic sample is available. Our test is detailed in this
section.

\subsection{Photometry for SED constructions}\label{sec:psf-match-phot}

    When constructing SEDs, photometry is usually done on images that have been 
PSF-matched across all bands, i.e., images taken in different bands are 
convolved so that each matches the band with the largest PSF. 
In our case, we convolved the images in the bluer bands to match the PSF size 
of the F444W image for each field. The PSFs used for this step were empirically 
constructed from isolated stars on a per-field, per-filter basis, following the 
procedures described in \citet[][]{LY2022}. We then generated the convolution 
kernels using the {\tt pypher} tool \citep[][]{Boucaud16_pypher} and applied 
them using {\tt astropy.convolution} \citep[][]{Astropy2022} to match the 
images in each band to the F444W image. Using the F444W image as the detection 
image, we then ran \texttt{SExtractor} again in dual-image mode and derived 
colours from the {\tt MAG\_ISO} magnitudes. 

    An additional step at this stage would be to rescale the 
{\tt MAG\_ISO}-based SED to a total-flux SED by adding the difference between 
the {\tt MAG\_ISO} and {\tt MAG\_AUTO} magnitudes in the F444W band. However, 
since many of our dropouts have nearby neighbours that contaminate the 
{\tt MAG\_AUTO} measurement, we chose not to apply this correction. 

    Because the F444W images are generally shallower than the F356W ones, 
a small fraction of our candidates could not be extracted when F444W was used 
as the detection band. For these sources, we reran \texttt{SExtractor} using 
the convolved F356W image as the detection image instead. However, a few 
sources with close neighbours still could not be extracted with this approach 
due to the larger PSF of the convolved F356W image that makes our target 
blended with the neighbours. For these remaining cases, we ran 
\texttt{SExtractor} once more, this time using the original (unconvolved) F356W 
image for detection while still using the convolved images for the photometric 
measurements. 

\subsection{SED fitting Purification Schemes}

    Following \citet[][]{Yan_highz_23}, we performed SED fitting using two 
frequently used but very different tools, 
Le Phare \citep[][]{Arnouts1999,Ilbert2006} 
and EAZY-py \citep[version 0.8.2;][]{Brammer2008}. The sample purification 
schemes based on the results from Le Phare and EAZY-py are referred to as 
schemes ``S1'' and ``S2'', respectively. 

    In S1, we constructed templates using the population synthesis model of
BC03 \citep[][]{Bruzual2003}, assuming a Chabrier initial mass function
\citep[][]{Chabrier2003} and an exponentially declining star formation history 
in the form of SFR~$\propto e^{-t/\tau}$, with $\tau$ ranging from 0 to 13 Gyr. 
We adopt the Calzetti extinction law, with $E(B-V)$ ranging from 0 to 1~mag in 
steps of 0.1~mag. Contributions from emission lines in individual filters are 
included by enabling both {\tt EM\_LINES} and {\tt ADD\_EMLINES} options 
in the Le Phare configuration file. 

   In S2, we adopt two sets of templates from \citet[][]{Larson2023_highz-templates}, 
hereafter referred to as ``sets 3 and 4'' (set~3+4), which are specifically 
optimised for fitting high-redshift galaxies. These templates have the 
Ly$\alpha$ emission reduced or omitted (10\% in set~3 and 0\% in set~4). 

   In both schemes, 0.05~mag was added in quadrature to the photometric errors 
to account for systematic errors, and $2\sigma$ detection upper limits were 
used in bands with non-detections. 

\begin{figure*}
    \centering
    \includegraphics[width=0.9\linewidth]{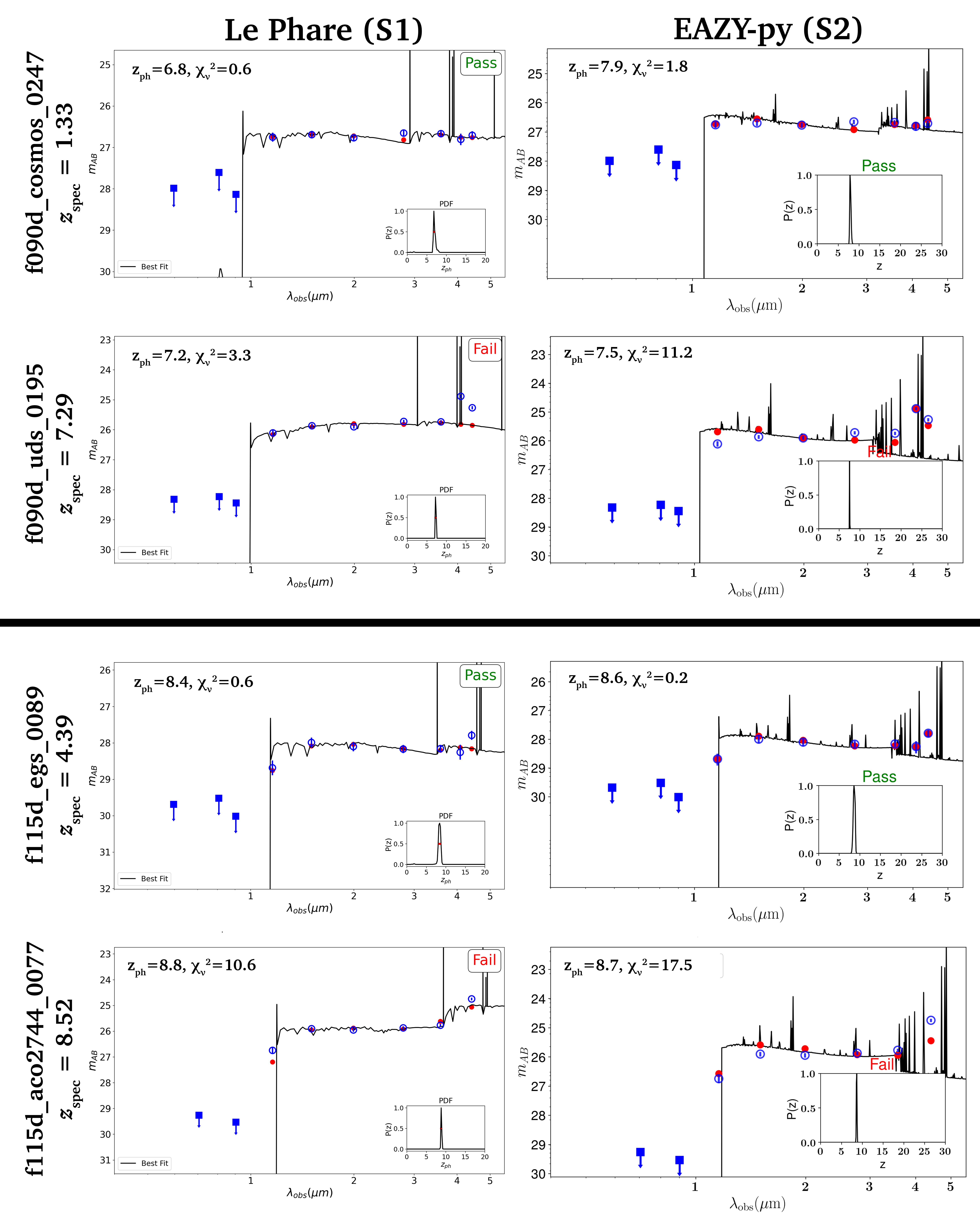}
    \caption{
    Examples of contamination and overkill cases in the SED-fitting-based 
    sample purification schemes. The left and right panels show the best-fit 
    SEDs from Le Phare (used in S1) and EAZY-py (used in S2), respectively; 
    the inset panels show the corresponding photometric redshift ($z_{\rm ph}$) 
    probability distribution functions (PDF). 
    In all plots, the photometric data points and the upper 
    limits that construct the SED are based on photometry described in 
    Appendix~\ref{sec:psf-match-phot}. In this illustration, a 
    reduced-$\chi^2$ threshold of $\chi_\nu^2=3$ is adopted, and the dropouts 
    below or above this threshold are marked as ``Pass'' or ``Fail'',
    respectively.
    The upper two rows show two F090W dropouts, one being an interloper at 
    $z_{\rm spec}=1.33$ and the other being a confirmed galaxy at 
    $z_{\rm spec}=7.29$, 
    while the lower two rows show two F115W dropouts, one being an
    interloper at $z_{\rm spec}=4.39$ and the other being a confirmed galaxy at $z_{\rm spec}=8.52$. 
    In both interloper cases, they have good high-$z$ solutions 
    with $\chi_\nu^2\leq3$ and narrow PDF in both S1 and S2 and would therefore be retained, i.e., they are cases of contamination.
    On the other hand, the two confirmed high-$z$ galaxies have $\chi_\nu^2>3$ 
    and would be rejected by both schemes, i.e., they are cases of overkill.
    }
    \label{fig:sed_examples}
\end{figure*}

\subsection{Performance of Purification Schemes}\label{appendix:zph-vs-zspec}

   The contaminant rejection in S1 and S2 was done in two steps. First,
the objects that could not be well fitted were rejected. This was done by
removing any objects that have the reduced fitting $\chi^2$ values (hereafter 
denoted as $\chi^2_\nu$) about a certain threshold. For this test, we varied 
the threshold over a large range. Second, as in Section~\ref{sec:specz}, any 
objects with $z_{\rm phot}<6.4$ ($z_{\rm phot}<8.4$) were deemed as 
contaminants to the F090W-dropout (F115W-dropout) sample and were rejected.

   We evaluate the performance of S1 and S2 using the spectroscopic sample.
In each scheme, there are contaminants being retained and bona fide high-$z$
objects being rejected, either of which is undesirable. 
Figure~\ref{fig:sed_examples} shows some examples. 

For a given reduced-$\chi^2$ (hereafter $\chi^2_\nu$) threshold, we define the 
``overkill rate'' as 
$f_{\rm overkill} = N_{\rm high\text{-}{\it z},rejected} / N_{\rm high\text{-}{\it z},spec}$, 
where $N_{\rm high\text{-}{\it z},spec}$ is the number of spectroscopically 
confirmed high-$z$ galaxies and $N_{\rm high\text{-}{\it z},rejected}$ is 
its subset that is rejected by the purification scheme. 
We also define the contamination rate as 
$f_{\rm ctm,SED} = N_{\rm interloper,retained} / N_{\rm spec,retained}$ 
where $N_{\rm interloper,retained}$ is the number of actual low-$z$ galaxies that were retained as high-$z$ ones by the purification scheme, 
and $N_{\rm spec,retained}$ is the number of all spectroscopically identified objects retained in that scheme. 
A useful purification scheme should have a small $f_{\rm ctm,SED}$ without leading to a large $f_{\rm overkill}$. 

\begin{figure*}
    \includegraphics[width=0.8\textwidth]{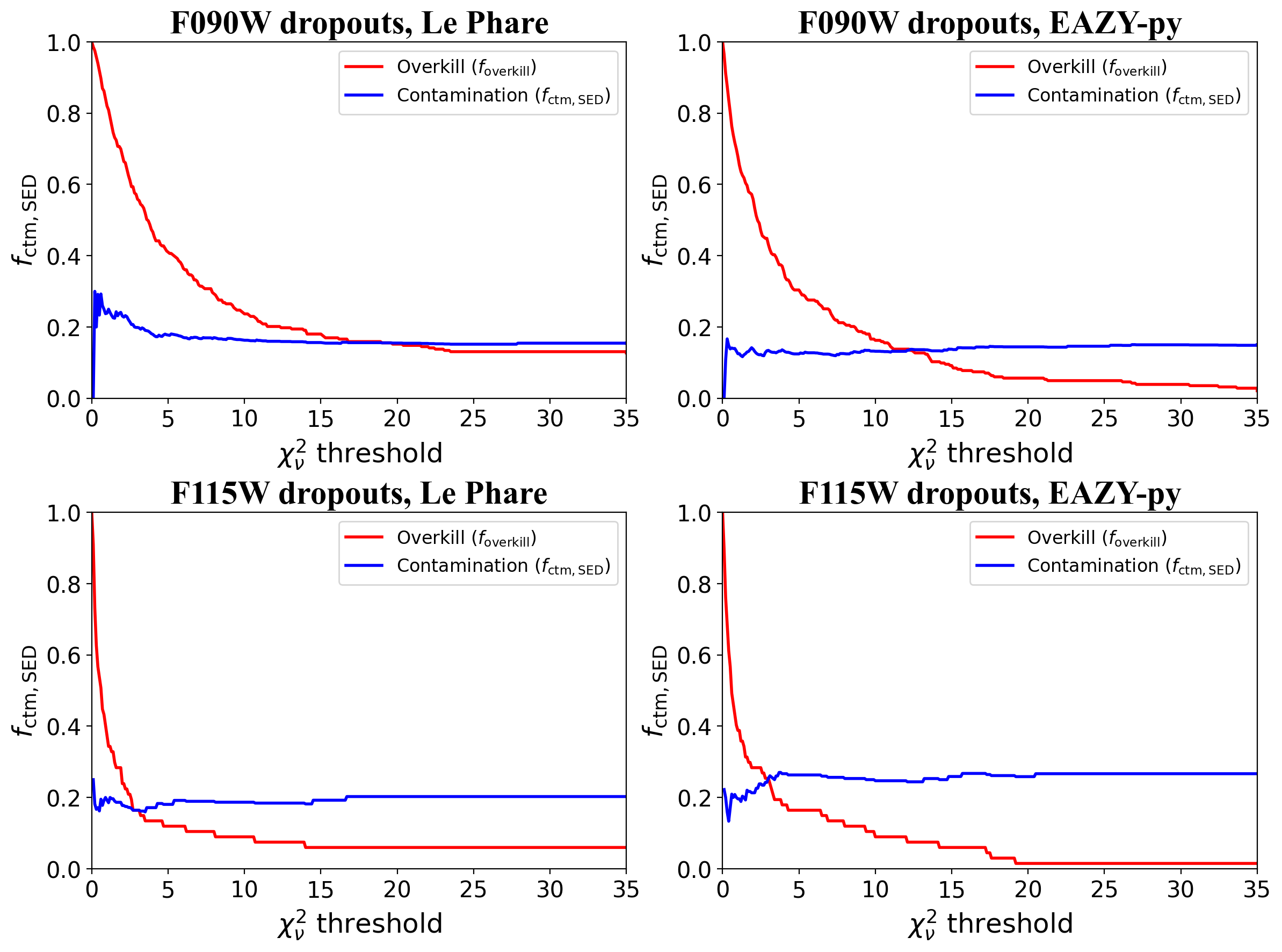}
    \label{fig:sed-overkill}
    \caption{
    Performance of the SED-fitting-based sample purification evaluated using 
    the spectroscopic sample. 
    The upper and lower rows show the F090W and F115W dropout samples, respectively; 
    the left and right columns show results from Le Phare (S1) and EAZY-py (S2), respectively. 
    The blue curves give the contamination rate, and the red curves show the overkill rate. 
    }
\end{figure*}

   Figure~\ref{fig:sed-overkill} shows $f_{\rm overkill}$ and $f_{\rm ctm,SED}$ 
for the two dropout samples in both purification schemes. The most striking
point is that $f_{\rm overkill}$ remains high until a large $\chi^2_\nu$ is
adopted. For example, at the threshold of $\chi^2_\nu =3$, $f_{\rm overkill}$ 
is 55.8\% and 43.1\% in S1 and S2 for the F090W dropouts, respectively. The 
situation is better for the F115W dropouts, but $f_{\rm overkill}$ is still
16.4\% and 25.4\% in S1 and S2, respectively. Even if the threshold is increased
to $\chi^2_\nu =10$, $f_{\rm overkill}$ remains high at 23.5\% and 16.3\% in
S1 and S2 for the F090W dropouts, respectively, although it does lower to
9\% in both S1 and S2 for the F115W dropouts. On the other hand,
$f_{\rm ctm,SED}$ keeps a rather flat trend with increasing $\chi^2_\nu$
threshold; after the threshold being increased to 3 and higher, 
$f_{\rm ctm,SED}$ is $\sim$18\% and $\sim$13\% in S1 and S2 for the F090W
dropouts, respectively, and $\sim$18\% and $\sim$25\% in S1 and S2 for the
F115W dropouts, respectively. Ironically, decreasing the threshold to below 3
will not necessarily decrease $f_{\rm ctm,SED}$, because more genuine 
high-$z$ galaxies will be overkilled and therefore the denominator becomes
smaller. 

   The root cause of the problems revealed in this test is that the templates
in use do not yet represent real galaxies well enough, especially those for
high-$z$ galaxies. Any purification schemes on dropout samples, or any
methods relying on the direct use of $z_{\rm phot}$ to select high-$z$ galaxies
samples, should be applied with caution.

\bsp	
\label{lastpage}
\end{document}